\documentclass[twocolumn,preprintnumbers,amsmath,amssymb]{revtex4-1}

\usepackage[utf8]{inputenc} 
\usepackage{setspace}

\usepackage{amstext,amssymb,amsfonts}
\usepackage{bm}
\usepackage{amsmath}

\usepackage{graphicx}
\usepackage{graphics}
\usepackage{epsfig}
\usepackage{graphicx,latexsym,amsmath,amssymb,amsfonts}

\usepackage{txfonts}

\usepackage{hyperref}

\usepackage{amstext,amsmath,amssymb,amsfonts}
\usepackage{bm}

\usepackage{mathtools,slashed}

\usepackage{xcolor}
\usepackage{color}
\definecolor{lightblue}{rgb}{0.2,0.2,0.7}
\definecolor{darkblue}{rgb}{0,0.25,0.5}
\definecolor{redbrown}{rgb}{0.875,0.25,0.125}
\definecolor{darkgreen}{rgb}{0,0.5,0}

\newcommand{\bra}[1]{\ensuremath{\langle #1 \vert}}
\newcommand{\ket}[1]{\ensuremath{\vert #1  \rangle}}

\renewcommand{\b}[1]{\ensuremath{\mathbf{#1}}}

\renewcommand{\d}{\ensuremath{\text{d}}}

\newcommand{\lr}{\ensuremath{\text{lr}}}
\newcommand{\sr}{\ensuremath{\text{sr}}}
\newcommand{\ee}{\ensuremath{\text{ee}}}

\DeclareMathOperator{\erf}{erf}
\DeclareMathOperator{\erfc}{erfc}

\newcommand{\ct}{\ensuremath{\tilde{c}}}

\begin{document}
\title{Four-component relativistic range-separated density-functional theory:\\ Short-range exchange local-density approximation}
\author{Julien Paquier} \email{julien.paquier@lct.jussieu.fr}
\author{Julien Toulouse} \email{toulouse@lct.jussieu.fr}
\affiliation{Laboratoire de Chimie Th\'eorique (LCT), Sorbonne Universit\'e and CNRS, F-75005 Paris, France}
\date{Octobre 19, 2018}
\begin{abstract}
We lay out the extension of range-separated density-functional theory to a four-component relativistic framework using a Dirac-Coulomb-Breit Hamiltonian in the no-pair approximation. This formalism combines a wave-function method for the long-range part of the electron-electron interaction with a density(-current) functional for the short-range part of the interaction. We construct for this formalism a short-range exchange local-density approximation based on calculations on a relativistic homogeneous electron gas with a modified Coulomb-Breit electron-electron interaction. More specifically, we provide the relativistic short-range Coulomb and Breit exchange energies per particle of the relativistic homogeneous electron gas in the form of Pad\'e approximants which are systematically improvable to arbitrary accuracy. These quantities, as well as the associated effective Coulomb-Breit exchange hole, show the important impact of relativity on short-range exchange effects for high densities. 
\end{abstract}

\maketitle

\section{Introduction}

Range-separated density-functional theory (RS-DFT) (see, e.g., Refs.~\onlinecite{Sav-INC-96,TouColSav-PRA-04}) is an alternative to Kohn-Sham density-functional theory (DFT)~\cite{KohSha-PR-65} for electronic-structure calculations of atoms, molecules, and solids. It permits to rigorously combine an explicit wave-function calculation for the long-range part of the electron-electron interaction with a compact density functional for the complement short-range part of the electron-electron interaction. RS-DFT leads to a faster basis convergence than standard wave-function methods~\cite{FraMusLupTou-JCP-15} and can provide improvement over usual Kohn-Sham DFT approximations for the description of strong-correlation effects (see, e.g., Refs.~\onlinecite{PolSavLeiSto-JCP-02,FroTouJen-JCP-07}), weak intermolecular interactions (see, e.g., Refs.~\onlinecite{AngGerSavTou-PRA-05,GolWerSto-PCCP-05,TouGerJanSavAng-PRL-09}), fractionally charged subsystems (see, e.g., Refs.~\onlinecite{VydScuPer-JCP-07,MusTou-MP-17}), and electronic excitation energies (see, e.g., Refs.~\onlinecite{TawTsuYanYanHir-JCP-04,RebSavTou-MP-13,FroKneJen-JCP-13}).

For the description of compounds with heavy elements, relativistic effects have to be incorporated into RS-DFT. A simple approach that has been used consists in using standard scalar-relativistic effective-core potentials in RS-DFT~\cite{FroReaWahWahJen-JCP-09,AlaFro-CPL-12}. A more sophisticated approach was developed by Kullie and Saue~\cite{KulSau-CP-12} who extended RS-DFT to a four-component relativistic Dirac-Coulomb Hamiltonian, using second-order M{\o}ller-Plesset (MP2) perturbation theory for the long-range part of the calculation and usual non-relativistic short-range semi-local exchange-correlation density-functional approximations. To make this approach more rigorous and possibly more accurate, especially for core properties, relativistic short-range density-functional approximations should be used. A number of relativistic density-functional approximations have been proposed for four-component relativistic Kohn-Sham DFT (see, e.g., reviews in Refs.~\onlinecite{Eng-INC-02,Wul-INC-10}), but no relativistic short-range density-functional approximations has been developed yet. This is unfortunate since relativistic effects are most important in spatial regions of high density whose contribution to the energy in RS-DFT mainly comes from the short-range exchange-correlation density functional. 

In the present work, we remedy to this limitation by constructing a relativistic short-range exchange local-density approximation (LDA) based on a relativistic homogeneous electron gas (RHEG) with a modified electron-electron interaction. The choice to focus on the exchange energy and not on the correlation energy is motivated by the fact that, at high densities where relativistic effects are important, exchange largely dominates correlation, at least in not very strongly correlated systems. The choice to target a LDA, as opposed to a generalized-gradient approximation (GGA), is motivated by the fact that LDA is the standard first-level approximation to consider in DFT. Moreover, even though corrections to the relativistic LDA were shown to be important for the case of the full-range electron-electron interaction~\cite{EngKelFacMulDre-PRA-95,EngKelDre-PRA-96}, such corrections beyond LDA are expected to be much smaller for the case of a short-range electron-electron interaction, as known in non-relativistic RS-DFT~\cite{TouColSav-PRA-04,TouColSav-JCP-05}. Beyond the goal of using this relativistic short-range exchange LDA in RS-DFT calculations of molecular and solid-state systems, the present work also aims at analyzing the importance of relativistic effects on the short-range exchange energy.

The paper is organized as follows. In Section~\ref{sec:rsdft}, we briefly lay out the formalism of RS-DFT for a four-component relativistic Dirac-Coulomb-Breit Hamiltonian. In Sections~\ref{sec:rhegI},~\ref{sec:rhegII}, and~\ref{sec:rhegIII}, we review the calculation of the full-range exchange energy per particle of the RHEG with the standard Dirac-Coulomb-Breit Hamiltonian in a way that will prepare for the extension to the short-range case. We discuss the importance of the separate Coulomb and Breit contributions to the exchange energy per particle and to the exchange hole. In Section~\ref{sec:srrheg}, we derive the exchange energy per particle of a RHEG with the short-range version of the Coulomb-Breit electron-electron interaction. Whereas the exchange energy per particle of a non-relativistic homogeneous electron gas with a short-range interaction can be obtained analytically quite easily~\cite{Sav-INC-96,GilAdaPop-MP-96,TouSavFla-IJQC-04}, the calculation of the relativistic analogue turned out to be quite a formidable task. We did not manage to obtain a closed-form expression for the relativistic short-range exchange energy per particle, but we could express it as a divergent series in inverse powers of the speed of light that can be summed to high accuracy using Pad\'e approximants, which is suitable for all practical purposes. In Section~\ref{sec:expansions}, we derive expansions of the relativistic short-range exchange energy per particle for small and large values of the range-separation parameter, and use them to construct a simple approximation to the relativistic short-range exchange energy per particle which can be used as an alternative to the Pad\'e approximants. Finally, Section~\ref{sec:conclusions} contains our conclusions and prospects for future work. Details of the calculations and the complete expressions obtained are given in the Supplementary Information~\cite{PaqTou-JJJ-XX-note}.

\section{Relativistic range-separated density-functional theory}
\label{sec:rsdft}

\subsection{No-pair Dirac-Coulomb-Breit wave-function theory}

For relativistic electronic-structure calculations, a good starting point is the Dirac-Coulomb-Breit (DCB) electronic Hamiltonian in the no-pair approximation~\cite{Suc-PRA-80}
\begin{eqnarray}
\hat{H}^{\text{DCB}}_{+} &=& \hat{H}^{\text{D}}_{+} + \hat{W}_{+}^{\text{CB}},
\end{eqnarray}
where $\hat{H}^{\text{D}}_{+}$ is the one-electron Dirac Hamiltonian and $\hat{W}_{+}^{\text{CB}}$ is the two-electron Coulomb-Breit (CB) interaction. In second-quantized position representation, $\hat{H}^{\text{D}}_{+}$ can be written as, in atomic units, 
\begin{eqnarray}
\hat{H}^{\text{D}}_{+} &=&   \int \! \d\b{r} \; \hat{\psi}^\dagger_{+}(\b{r}) \left[ c \; (\bm{\alpha} \cdot \b{p}) +\bm{\beta} \; mc^{2} + v_{\text{ne}}(\b{r}) \b{I}_4 \right] \hat{\psi}_{+}(\b{r}), \;\;\;\;\;
\label{HD+}
\end{eqnarray}
where $\b{p} = -i \bm{\nabla}_{\b{r}}$ is the momentum operator, $c = 137.036$ a.u. is the speed of light, and $m=1$ a.u. is the electron mass that has been kept for clarity, $v_{\text{ne}}(\b{r})$ is the electron-nuclei scalar potential energy operator, $\b{I}_4$ is the $4 \times 4$ identity matrix, and $\bm{\alpha}$ and $\bm{\beta}$ are the $4 \times 4$ Dirac matrices
\begin{eqnarray}
\bm{\alpha} = \left(\begin{array}{cc}
\b{0}_2&\bm{\sigma}\\
\bm{\sigma}&\b{0}_2\\
\end{array}\right)
~\text{and}~~
\bm{\beta} = \left(\begin{array}{cc}
\b{I}_{2}&\b{0}_2\\
\b{0}_2&-\b{I}_{2}\\
\end{array}\right),
\end{eqnarray}
where $\bm{\sigma}=(\bm{\sigma}_x,\bm{\sigma}_y,\bm{\sigma}_z)$ is the 3-dimensional vector of the $2 \times 2$ Pauli matrices, and $\b{0}_2$ and $\b{I}_2$ are the $2 \times 2$ zero and identity matrices, respectively. In Eq.~(\ref{HD+}), $\hat{\psi}_{+}(\b{r})$ and $\hat{\psi}_{+}^\dagger(\b{r})$ are the (projected) annihilation and creation field operators
\begin{eqnarray}
\hat{\psi}_{+}(\b{r}) = \sum_p^+ \psi_p(\b{r}) \; \hat{a}_p ~~\text{and}~~ \hat{\psi}_{+}^\dagger(\b{r}) = \sum_p^+ \psi_p^\dagger(\b{r}) \; \hat{a}_p^\dagger,
\label{fieldop}
\end{eqnarray}
where the sum is over a set of orthonormal 4-component-spinor orbitals $\{\psi_p(\b{r})\}$ which are positive-energy eigenfunctions of a Dirac Hamiltonian with some potential~\cite{Mit-PRA-81}, and $\hat{a}_p$ and $\hat{a}_p^\dagger$ are the corresponding annihilation and creation operators of these orbitals. The restriction in the sum to positive-energy states corresponds to the no-pair approximation, in which negative-energy states corresponding to positronic states are disregarded. The two-electron Coulomb-Breit interaction term is written as
\begin{eqnarray}
\hat{W}^{\text{CB}}_{+} &=& \frac{1}{2} \iint \! \d\b{r}_1 \d\b{r}_2 \; \hat{\psi}^\dagger_{+}(\b{r}_1) \hat{\psi}^\dagger_{+}(\b{r}_2) \b{w}^{\text{CB}}(\b{r}_{12}) \hat{\psi}_{+}(\b{r}_2) \hat{\psi}_{+}(\b{r}_1), \;\;\;\;
\label{WCB+}
\end{eqnarray}
where $\b{w}^{\text{CB}}(\b{r}_{12})$ is the sum of the Coulomb and Breit contributions
\begin{eqnarray}
\b{w}^{\text{CB}}(\b{r}_{12}) =  \b{w}^{\text{C}}(r_{12}) + \b{w}^{\text{B}}(\b{r}_{12}),
\end{eqnarray}
where $\b{r}_{12} = \b{r}_{1}-\b{r}_{2}$ and $r_{12} = |\b{r}_{12}|$. The Coulomb interaction is
\begin{eqnarray}
\b{w}^{\text{C}}(r_{12}) = w_\ee(r_{12}) \; (\b{I}_{4})_{1} \; (\b{I}_4)_2,
\label{wC}
\end{eqnarray}
where $w_\ee(r_{12}) = 1/r_{12}$, and $(\b{I}_{4})_{1}$ and $(\b{I}_4)_2$ are the $4 \times 4$ identity matrices acting on electron 1 and 2, respectively. The Breit interaction is
\begin{eqnarray}
 \b{w}^{\text{B}}(\b{r}_{12}) = -\frac{1}{2} w_\ee(r_{12}) \left( \bm{\alpha}_{1} \cdot \bm{\alpha}_{2} + \frac{(\bm{\alpha}_{1}\cdot\b{r}_{12}) ~ (\bm{\alpha}_{2}\cdot\b{r}_{12})}{r_{12}^{2}} \right), 
\label{wB}
\end{eqnarray}
where the Dirac matrices $\bm{\alpha}_{1}$ and $\bm{\alpha}_{2}$ act on electron 1 and 2, respectively. The Coulomb-Breit interaction corresponds to the single-photon exchange electron-electron scattering amplitude in quantum-electrodynamics (QED) evaluated with the zero-frequency limit of the photon propagator in the Coulomb electromagnetic gauge. More specifically, the instantaneous Coulomb interaction corresponds to the longitudinal component of the photon propagator, whereas the Breit interaction is obtained from to the zero-frequency transverse component of the photon propagator. The Breit interaction comprises the instantaneous magnetic Gaunt interaction, $-w_\ee(r_{12}) \bm{\alpha}_{1} \cdot \bm{\alpha}_{2}$, and the remaining lowest-order retardation correction (see, e.g., Ref.~\onlinecite{GorInd-PRA-88}).

The no-pair Hamiltonian is not unique since it depends on the choice of the set of orbitals $\{\psi_p\}$. It has been proposed~\cite{SauVis-INC-03,AlmKneJenDyaSau-JCP-16} to define the no-pair relativistic ground-state energy of a $N$-electron system using a minmax principle, that we will formally write as
\begin{eqnarray}
E  &=& \min_{\Psi} \left[ \max_{\{\psi_p\}} \bra{\Psi} \hat{H}^{D}_+ + \hat{W}_{+}^{\text{CB}} \ket{\Psi} \right],
\label{EWFT}
\end{eqnarray}
where the maximization is done with respect to the set of positive-energy orbitals $\{\psi_p\}$ [on which the Hamiltonian is projected via Eq.~(\ref{fieldop})] by rotations with its complement set of negative-energy orbitals, and the minimization is done with respect to normalized multideterminant wave functions $\Psi$ within the $N$-electron space generated by the set of positive-energy orbitals $\{\psi_p\}$. In practice, Eq.~(\ref{EWFT}) can be realized with a multiconfiguration self-consistent-field (MCSCF) algorithm which targets a saddle point in the parameter space~\cite{JenDyaSauFae-JCP-96,ThyFleJen-JCP-08,AlmKneJenDyaSau-JCP-16}. For one-electron Dirac Hamiltonians, this type of minmax principle appears well founded (see Refs.~\onlinecite{Tal-PRL-86,DatDev-PJP-88,GriSie-JLMS-99,DolEstSer-JFA-00}). For the no-pair Dirac-Coulomb-Breit Hamiltonian with a correlated wave function, to the best of our knowledge this minmax principle has not been rigorously examined mathematically but the results of Ref.~\onlinecite{AlmKneJenDyaSau-JCP-16} on two-electron systems suggests that it is indeed a reasonable definition of the no-pair ground-state energy.

\subsection{No-pair Dirac-Coulomb-Breit range-separated density-functional theory}

The Dirac-Coulomb RS-DFT introduced by Kullie and Saue~\cite{KulSau-CP-12} can be readily extended to a Dirac-Coulomb-Breit Hamiltonian. We will do so using the general formalism of relativistic current-density-functional theory~\cite{RajCal-PRB-73,Raj-JPC-78,MacVos-JPC-79}, even though we do not consider any external vector potential. 

The starting point is the following decomposition of the Coulomb-form electron-electron potential $w_\ee(r_{12}) = 1/r_{12}$ appearing in Eqs.~(\ref{wC}) and (\ref{wB})
\begin{eqnarray}
w_\ee(r_{12}) = w^{\lr,\mu}_\ee(r_{12}) + w^{\sr,\mu}_\ee(r_{12}),
\end{eqnarray}
where $w^{\lr,\mu}_\ee(r_{12})=\erf(\mu r_{12})/r_{12}$ and $w^{\sr,\mu}_\ee(r_{12})=\erfc(\mu r_{12})/r_{12}$ are long-range and short-range potentials, respectively. Here, $\erf(x)= (2/\sqrt{\pi}) \int_{0}^{x} e^{-t^{2}} \d t$ is the error function,  $\erfc(x) = 1 -\erf(x)$ is the complementary error function, and $\mu$ is the range-separation parameter acting like an inverse smooth cut-off radius. We then assume that the no-pair relativistic ground-state energy can be expressed as
\begin{eqnarray}
E  &=& \min_{\Psi} \left[ \max_{\{\psi_p\}} \left\{ 
\bra{\Psi} \hat{H}^{D}_{+} + \hat{W}_{+}^{\text{CB,lr},{\mu}} \ket{\Psi} + \bar{E}_{\text{Hxc}}^{\text{CB,sr},{\mu}}[n_{\Psi},\b{j}_{\Psi}] 
\right\} \right], \;\;\;\;\;\;
\label{ERSDFT}
\end{eqnarray}
where, as before, the maximization is over the set of projecting positive-energy orbitals $\{\psi_p\}$ by rotations with its complement set of negative-energy orbitals and the minimization is over normalized multideterminant wave functions $\Psi$ within the $N$-electron space generated by the set of positive-energy orbitals. In Eq.~(\ref{ERSDFT}), $\hat{W}_{+}^{\text{CB,lr},{\mu}}$ is the long-range version of the two-electron Coulomb-Breit interaction as defined by Eqs.~(\ref{WCB+})-(\ref{wB}) but with the substitution $w_\ee(r_{12}) \longrightarrow w^{\lr,\mu}_\ee(r_{12})$, and $\bar{E}_{\text{Hxc}}^{\text{CB,sr},{\mu}}[n_{\Psi},\b{j}_{\Psi}]$ is a complement short-range Coulomb-Breit Hartree-exchange-correlation functional of the electron density $n_\Psi(\b{r}) = \bra{\Psi} \hat{n}(\b{r}) \ket{\Psi}$ and current electron density $\b{j}_\Psi(\b{r}) = \bra{\Psi} \hat{\b{j}}(\b{r}) \ket{\Psi}$, where $\hat{n}(\b{r}) = \hat{\psi}^\dagger_{+}(\b{r}) \hat{\psi}_{+}(\b{r})$ and $\hat{\b{j}}(\b{r}) = \hat{\psi}^\dagger_{+}(\b{r}) c \bm{\alpha} \hat{\psi}_{+}(\b{r})$ are the density and current density operators, respectively. The term ``complement short-range'' means that the functional contains not only a pure short-range contribution but also a mixed long-range/short-range contribution~\cite{TouSav-JMS-06}. We note that, even though Eq.~(\ref{ERSDFT}) seems a natural extension of non-relativistic RS-DFT~\cite{Sav-INC-96,TouColSav-PRA-04}, the existence of a universal density-current functional $\bar{E}_{\text{Hxc}}^{\text{CB,sr},{\mu}}[n,\b{j}]$ giving the no-pair relativistic ground-state energy via the minmax procedure of Eq.~(\ref{ERSDFT}) is not established. Indeed, similarly to the problem of defining static density functionals for excited-state energies (see, e.g., Refs.~\onlinecite{Gor-PRA-99,AyeLev-PRA-09}), it seems \textit{a priori} only possible to define a no-pair relativistic functional which either is universal but satisfies only a stationary principle (instead of a minmax principle) or satisfies a minmax principle but is not universal (i.e., depending on an external potential via the projecting orbitals $\{\psi_p\}$). It might be necessary to go at the QED level to rigorously formulate a RS-DFT approach, as it was done for Kohn-Sham DFT~\cite{Eng-INC-02}. These aspects are beyond the scope of the present work.

The limiting cases of the relativistic RS-DFT approach of Eq.~(\ref{ERSDFT}) should be mentioned. For $\mu\to\infty$, the long-range interaction reduces to the Coulomb form, $w^{\lr,\mu\to\infty}_\ee(r_{12})=1/r_{12}$, and the complement short-range functional vanishes, $\bar{E}_{\text{Hxc}}^{\text{CB,sr},{\mu\to\infty}}[n]=0$, so Eq.~(\ref{ERSDFT}) reduces to the wave-function theory of Eq.~(\ref{EWFT}). For $\mu=0$, the long-range interaction vanishes, $w^{\lr,\mu=0}_\ee(r_{12})=0$, and the complement short-range functional reduces to a full-range density functional, $\bar{E}_{\text{Hxc}}^{\text{CB,sr},{\mu=0}}[n]=E_{\text{Hxc}}^{\text{CB}}[n]$, so it reduces to a no-pair relativistic Kohn-Sham DFT method (see, e.g., Ref.~\onlinecite{SauHel-JCC-02}), and the corresponding minimizing wave function in Eq.~(\ref{ERSDFT}) is a single Slater determinant $\Psi^{\mu=0}=\Phi$.

We decompose now the short-range density-current functional into Hartree, exchange, and correlation contributions
\begin{eqnarray}
\bar{E}_{\text{Hxc}}^{\text{CB,sr},{\mu}}[n,\b{j}] &=& E_{\text{H}}^{\text{CB,sr},{\mu}}[n,\b{j}] + E_{\text{x}}^{\text{CB,sr},{\mu}}[n,\b{j}] + \bar{E}_{\text{c}}^{\text{CB,sr},{\mu}}[n,\b{j}].
\nonumber\\
\label{EHxcsrdecomp}
\end{eqnarray} 
The short-range Hartree contribution, containing Coulomb and Breit contributions, is an explicit functional of the density $n$ and the current $\b{j}$
\begin{multline}
E_{\text{H}}^{\text{CB,sr},{\mu}}[n,\b{j}] = \frac{1}{2} \iint w_{\text{ee}}^{\text{sr},{\mu}} (r_{12})  ~ n(\b{r}_{1}) n(\b{r}_{2}) ~ \d\b{r}_{1} \d\b{r}_{2} ~~~~~~~~ 
\\
-\frac{1}{4c^2} \Bigg[ \iint w_{\text{ee}}^{\text{sr},{\mu}} (r_{12}) ~{\b{j}(\b{r}_{1})\cdot \b{j}(\b{r}_{2})} \; \d\b{r}_{1} \d\b{r}_{2} ~~~~~~  
\\
+ \iint w_{\text{ee}}^{\text{sr},{\mu}} (r_{12}) ~ \frac{\b{j}(\b{r}_{1})\cdot\b{r}_{12} ~~~ \b{j}(\b{r}_{2})\cdot\b{r}_{12}}{r_{12}^{2}} \d\b{r}_{1} \d\b{r}_{2} \Bigg].
\end{multline}
Note that, in relativistic Kohn-Sham DFT, the Hartree energy is usually defined with the full QED photon propagator in the Feynman gauge~\cite{Eng-INC-02}. Here, instead, we define the short-range Hartree energy with the two-electron Coulomb-Breit interaction in the Coulomb gauge, in order to be consistent with the corresponding long-range wave-function contribution. The short-range exchange density-current functional is consequently defined by
\begin{eqnarray}
E_{\text{x}}^{\text{CB,sr},{\mu}}[n,\b{j}] = \bra{\Phi[n,\b{j}]} \; \hat{W}_{+}^{\text{CB,sr},{\mu}} \; \ket{\Phi[n,\b{j}]} - E_{\text{H}}^{\text{CB,sr},{\mu}}[n,\b{j}],
\nonumber\\
\end{eqnarray}
where ${\Phi[n,\b{j}]}$ is the Kohn-Sham determinant associated with density $n$ and current $\b{j}$, and $\hat{W}_{+}^{\text{CB,sr},{\mu}}$ is the short-range version of the two-electron Coulomb-Breit interaction obtained as in Eqs.~(\ref{WCB+})-(\ref{wB}) but with the substitution $w_\ee(r_{12}) \longrightarrow w^{\sr,\mu}_\ee(r_{12})$. In Eq.~(\ref{EHxcsrdecomp}), $\bar{E}_{\text{c}}^{\text{CB,sr},{\mu}}[n,\b{j}]$ is the complement short-range correlation functional including all interaction effects beyond Hartree and exchange. Finally, note that, for closed-shell systems, the current vanishes, $\b{j}=\b{0}$, and thus we have simply short-range functionals of the density only. We only consider this case  in this work.

In practice, the application of the relativistic RS-DFT approach of Eq.~(\ref{ERSDFT}) requires to use approximations for the long-range wave-function part and for the short-range exchange and correlation functionals. For the long-range wave-function part, one can use standard approximations: Hartree-Fock, MP2, MCSCF, etc. For the short-range functionals, no approximation including the relativistic effects has been proposed so far. The rest of paper is devoted to the development of a relativistic LDA for the short-range exchange functional
\begin{eqnarray}
E_{\text{x,LDA}}^{\text{CB,sr},{\mu}}[n] = \int n(\b{r}) ~ {\epsilon}_{\text{x}}^{\text{CB,sr},{\mu}}\Big(n(\b{r})\Big) ~ \d\b{r},
\end{eqnarray}
where ${\epsilon}_{\text{x}}^{\text{CB,sr},{\mu}}(n)$ is the exchange energy per particle of a RHEG with the short-range two-electron Coulomb-Breit interaction. The calculation of the latter quantity is quite complicated and was performed with the help of the software Wolfram Mathematica \cite{Mathematica}.

\section{Exchange effects in the relativistic homogeneous electron gas}
\label{sec:rheg}

We consider a homogeneous electron gas, i.e. a box of volume $V$ containing $N$ electrons with electronic density $n=N/V$ characteristic of the system and a uniform background of positive charges with density $n_{\text{b}}$ equal to $n$ in order to ensure the electroneutrality of the system. The electron gas is studied in the thermodynamic limit, i.e. $N\rightarrow {\infty}$ and $V\rightarrow {\infty}$ while $n=N/V$ is kept constant, the positive background cancelling the diverging Hartree energy term. 
Such an electron gas is considered to be relativistic when the Fermi wave vector $k_{\text{F}} = (3 {\pi}^{2} n)^{1/3}$ is non negligible compared to the speed of light $c$. This must really be understood as a comparison between the energy related to the Fermi wave vector ${\hbar}k_{\text{F}}c$ and the rest energy $mc^{2}$ where we take these quantities in atomic units. 

In order to get an idea of the maximal value of $k_{\text{F}}$ that one encounters in an heavy element, one may consider a 1s hydrogen-like orbital, ${\psi}_{1\text{s}}(r) = \big(Z^{3}/{\pi}\big)^{1/2} e^{-Zr}$ where $Z$ is the atomic number, and calculate the density of this doubly occupied 1s orbital at the nucleus: $n = 2~|{\psi}_{1\text{s}}(0)|^{2} =  2~Z^{3}/{\pi}$ corresponding to a Fermi wave vector of $k_{\text{F}}= (6{\pi})^{1/3}~Z = 2.66~Z$. The heaviest elements having atomic numbers $Z$ of about 100, this corresponds to a maximal value of $k_\text{F}$ of nearly 300 a.u..

\subsection{One-electron part}
\label{sec:rhegI}

The form of the non-interacting one-electron Dirac equation for this homogeneous electron gas is
\begin{eqnarray}
\left(c(\bm{\alpha} \cdot \b{p})+\bm{\beta}mc^{2}\right){\psi}_{\b{k},{\sigma}}(\b{r})=E_{k} ~{\psi}_{\b{k},{\sigma}}(\b{r}),
\label{Diraconeelec}
\end{eqnarray}
where ${\psi}_{\b{k},{\sigma}}(\b{r})$ is a one-electron wave function with wave vector $\b{k}$ and ``spin'' index $\sigma=\uparrow,\downarrow$ associated with the positive-energy eigenvalue
\begin{eqnarray}
E_{k} =\sqrt{k^{2}c^{2}+m^{2}c^{4}}.
\end{eqnarray}
The wave functions ${\psi}_{\b{k},{\sigma}}(\b{r})$ are four-component spinors of the form
\begin{eqnarray}
{\psi}_{\b{k},{\sigma}}(\b{r}) &=& \left(\begin{array}{c}
{\varphi}_{\b{k},{\sigma}}(\b{r})\\
{\chi}_{\b{k},{\sigma}}(\b{r})\\
\end{array}\right),
\end{eqnarray}
where each component is itself a two-component spinor. The large component ${\varphi}_{\b{k},{\sigma}}(\b{r})$ has a plane-wave form
\begin{eqnarray}
{\varphi}_{\b{k},{\sigma}}(\b{r}) = \frac{1}{\sqrt{V}} \sqrt{\frac{E_{k}+mc^{2}}{2E_{k}}}e^{-i\b{k}\cdot\b{r}} {\varphi}_{\sigma},
\end{eqnarray}
and the small component is obtained from the large component by
\begin{eqnarray}
{\chi}_{\b{k},{\sigma}}(\b{r}) = \frac{c(\bm{\sigma}\cdot \b{k})}{E_{k}+mc^{2}}{\varphi}_{\b{k},{\sigma}}(\b{r}),
\end{eqnarray}
and ${\varphi}_{\sigma}$ is a two-component spin part with ${\small{\varphi}_{\uparrow}=\Big(\begin{array}{c}
1\\
0\\
\end{array}\Big)}$ and ${\small{\varphi}_{\downarrow}=\Big(\begin{array}{c}
0\\
1\\
\end{array}\Big)}$. What we call ``spin'' here really refers to the index $\sigma$ identifying the two components of the large-component spinor.

As in the non-relativistic case, we will calculate the exchange energy of the relativistic homogeneous electron gas using these non-interacting one-electron wave functions ${\psi}_{\b{k},{\sigma}}(\b{r})$. We note however that, in the relativistic case, there is in principle an ambiguity on which one-electron wave functions to use to calculate the exchange energy. Indeed, adding for example the Hartree-Fock exchange potential in Eq.~(\ref{Diraconeelec}) will \textit{a priori} lead to different one-electron wave functions since translational symmetry only imposes the real-space plane-wave form but not the spinor components. In the non-relativistic case, this ambiguity does not appear since in this case the one-electron plane-wave wave functions are completely determined by translational symmetry, independently of the one-electron potential used. We leave further discussions of this interacting point for future work.

\subsection{Full-range Coulomb-Breit exchange energy}
\label{sec:rhegII}

As the preparation for the case of the short-range Coulomb-Breit interaction, we recall the form of the full-range Coulomb-Breit exchange energy. The Coulomb exchange energy per particle can be expressed as~\cite{Jan-NC-62,MacVos-JPC-79} (see Appendix~\ref{app:rheg})
\begin{widetext}
\begin{eqnarray}
{\epsilon}_{\text{x}}^{\text{C}} &=&  -\frac{3~k_{\text{F}}}{4{\pi}}\Bigg(\frac{5}{6} + \frac{1}{3}{\ct}^{2} + \frac{2}{3} \sqrt{1+ {\ct}^{2}} ~ \text{arcsinh}\left(\frac{1}{\ct}\right) - \frac{1}{3}\Bigg(1+ {\ct}^{2}\Bigg)^{2}~\text{ln}\left(1+\frac{1}{{\ct}^{2}}\right) -\frac{1}{2}\left(\sqrt{1+ {\ct}^{2}} -{\ct}^{2} \text{arcsinh}\left(\frac{1}{\ct}\right)\right)^{2} ~\Bigg),
\label{epsxCtext}
\end{eqnarray}
where $\ct = {c}/{k_{\text{F}}}$. The Breit exchange energy per particle has a similar form~\onlinecite{RamRaj-PRA-82} (see Appendix~\ref{app:rheg})
\begin{equation}
{\epsilon}_{\text{x}}^{\text{B}} = \frac{3~k_{\text{F}}}{4{\pi}}\Bigg(1 - 2\Big(1+{\ct}^{2}\Big)\Bigg(1 - {\ct}^{2} \Bigg(-2~\text{ln}\left({\ct}\right) + \text{ln} \left( 1 + {\ct}^{2} \right) \Bigg)\Bigg) + 2\left(\sqrt{1+ {\ct}^{2}} -{\ct}^{2} \text{arcsinh}\left(\frac{1}{\ct}\right) \right)^{2} \Bigg).
\label{epsxBtext}
\end{equation}
\end{widetext}
The Breit exchange energy per particle is an approximation to the exchange energy per particle obtained with the transverse component of the full QED photon propagator~\cite{Jan-NC-62,Raj-JPC-78,MacVos-JPC-79}. The exchange energy per particle obtained with the full QED photon propagator has in fact a simpler expression than the Coulomb-Breit one, thanks to the cancellation of many terms between the Coulomb and transverse components,
\begin{eqnarray}
{\epsilon}_{\text{x}}^{\text{QED}} &=& -\frac{3~k_{F}}{4{\pi}}\Bigg(1-\frac{3}{2} \left(\sqrt{1+ {\ct}^{2}} -{\ct}^{2} \text{arcsinh}\left(\frac{1}{\ct}\right) \right)^{2} \Bigg).
\end{eqnarray}

It is interesting to consider the non-relativistic and ultra-relativistic limits of the Coulomb and Breit exchange energies. In the non-relativistic (NR) limit ${\ct} \to \infty$ (or equivalently the low-density limit), only the Coulomb contribution survives
\begin{eqnarray}
{\epsilon}_{\text{x}}^{\text{C,NR}} &=& -\frac{3~k_{F}}{4{\pi}}  ~~~\text{and}~~~ {\epsilon}_{\text{x}}^{\text{B,NR}} = 0,
\end{eqnarray}
which is consistent with the fact that the Breit interaction is a purely relativistic phenomenon. In the opposite ultra-relativistic (UR) limit ${\ct} \to 0$ (or equivalently the high-density limit), we have
\begin{eqnarray}
{\epsilon}_{\text{x}}^{\text{C,UR}} = -\frac{(1+\text{ln}4)~k_{F}}{4{\pi}} ~~~\text{and}~~~ {\epsilon}_{\text{x}}^{\text{B,UR}} = \frac{3~k_{\text{F}}}{4{\pi}},
\end{eqnarray}
i.e. the Breit contribution becomes the opposite of the non-relativistic Coulomb exchange energy, and is larger in absolute value than the Coulomb contribution, implying that the total exchange energy becomes positive.

\begin{figure}[t]
        \centering
        \includegraphics[width=6cm,angle=270]{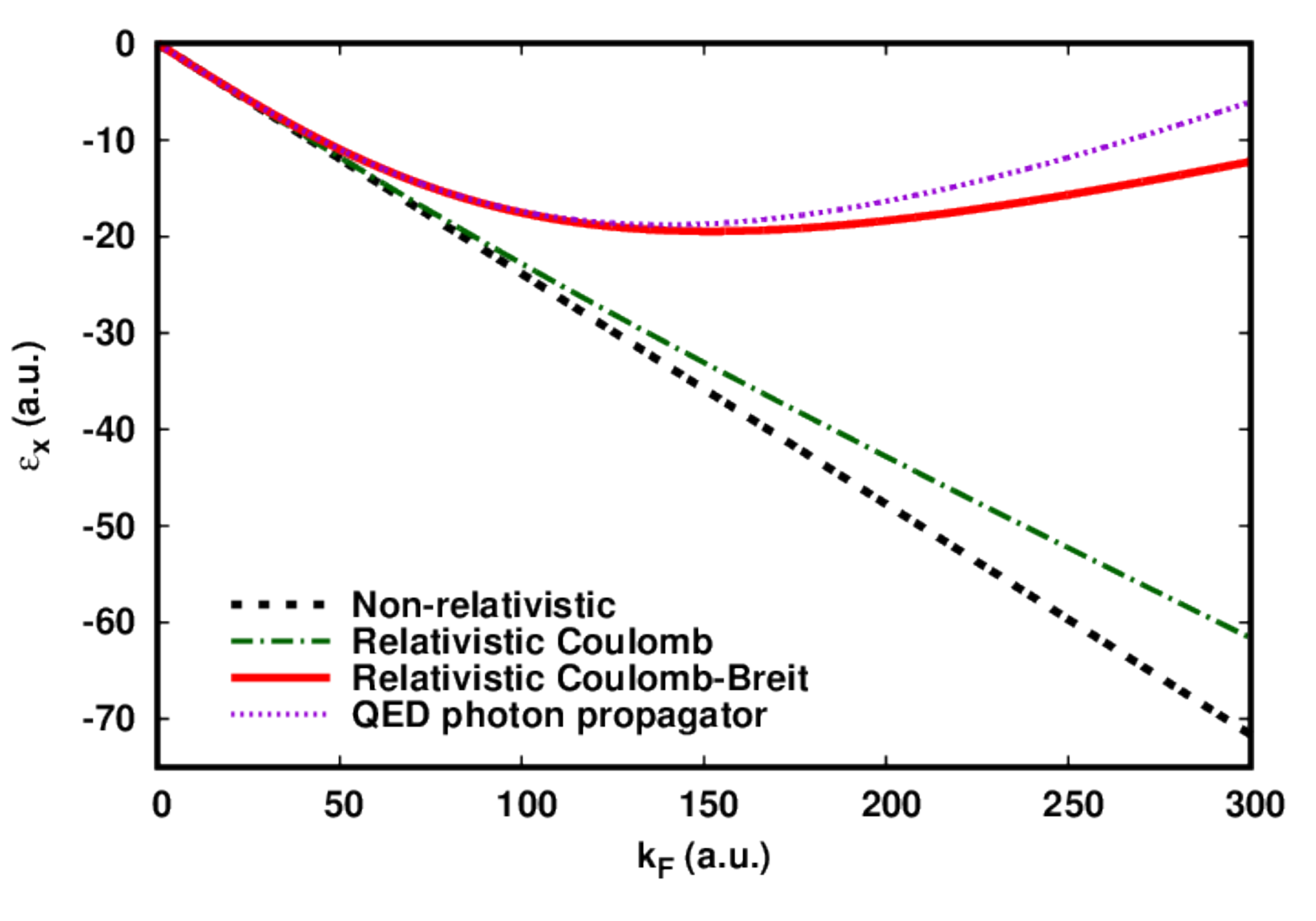}
\caption{Non-relativistic and relativistic exchange energies per particle for the Coulomb, Coulomb-Breit, and QED photon propagator electron-electron interactions as functions of the Fermi wave vector $k_\text{F}$.}
\label{fig:exchange}
\end{figure}

The different exchange energies per particle are plotted in Figure~\ref{fig:exchange} as functions of the Fermi wave vector $k_\text{F}$, up to $k_\text{F}=300$ a.u. which is about the largest value that could be encountered in an heavy element. We observe that, compared to the non-relativistic energy energy, relativity always reduces the exchange energy in absolute value, and this effect increases with the density. The relativistic effects for the Coulomb exchange contribution remain relatively small, even at high densities. By contrast, the Breit exchange contribution has a much larger effect at sufficiently high densities. The Coulomb-Breit exchange energy per particle is a good approximation to the exchange energy per particle obtained with the full QED photon propagator for $k_\text{F} \lesssim c \approx 137$ a.u., which is consistent with the fact that the Breit interaction constitutes only the leading term in the expansion of the QED photon propagator in $1/c$. 

Finally, we note that the exchange energy per particle corresponding to the Gaunt interaction can also be similarly obtained, but it starts to deviate from the transverse exchange energy obtained from the QED photon propagator for substantially smaller values of $k_\text{F}$ than the Breit exchange energy does. For this reason, we prefer in this work to use the Breit interaction.

\subsection{Effective Coulomb-Breit exchange hole}
\label{sec:rhegIII}

A convenient way of analyzing the exchange effects in terms of the interelectronic distance $r_{12}$ is to introduce an effective Coulomb-Breit exchange hole $n_{\text{x}}^\text{CB}(r_{12})$ such that
\begin{eqnarray}
	\epsilon_{\text{x}}^\text{CB} &=& \frac{1}{2} \int n_{\text{x}}^\text{CB}(r_{12}) ~ w_{\text{ee}}(r_{12}) ~ \d\b{r}_{12}.
\end{eqnarray}
For the case of the Coulomb interaction, the associated exchange hole was derived by Ellis~\cite{Ell-JPB-77} and MacDonald and Vosko~\cite{MacVos-JPC-79} (see Appendix~\ref{app:rheg})
\begin{multline}
	n_{\text{x}}^{\text{C}}(r_{12}) = - \frac{9}{4} n\frac{1}{(k_{\text{F}}r_{12})^{2}}\Bigg[j_{1}(k_{\text{F}} r_{12})^{2}+ (1-{\lambda})A_{\lambda}(k_{\text{F}} r_{12})^{2}
	\\
	+ {\lambda}B_{\lambda}(k_{\text{F}} r_{12})^{2}\Bigg],
\label{Coulombhole}
\end{multline}
with $\lambda=1/(1+\ct^2)$ and 
\begin{eqnarray}
	A_{\lambda}(k_{\text{F}} r_{12}) = \sum_{\nu=0}^{\infty} \frac{(2\nu+1)!!}{(2\nu+1)}~j_{\nu+1}(k_{\text{F}} r_{12})\left(\frac{\lambda}{k_{\text{F}} r_{12}} \right)^{\nu},
\nonumber\\
	B_{\lambda}(k_{\text{F}} r_{12}) = \sum_{\nu=0}^{\infty} \frac{(2\nu+1)!!}{(2\nu+1)}~j_{\nu+2}(k_{\text{F}} r_{12})\left(\frac{\lambda}{k_{\text{F}} r_{12}} \right)^{\nu},
\end{eqnarray}
where $j_{\nu}$ are the spherical Bessel functions. We have extended this result to the case of the Breit interaction. The associated exchange hole is (see Appendix~\ref{app:rheg})
\begin{multline}
	n_{\text{x}}^{\text{B}}(r_{12}) = -\frac{9}{2}n \frac{1}{(k_{\text{F}} r_{12})^{2}}\Bigg[-j_{1}(k_{\text{F}} r_{12})^{2} + (1-{\lambda}) A_{\lambda}(k_{\text{F}} r_{12})^{2}\Bigg].
\label{Breithole}
\end{multline}
To the best of our knowledge this expression had not been derived before. This Breit exchange hole should not be interpreted as a modification of the pair density, but simply as giving after integration the Breit exchange energy.

Like the non-relativistic exchange hole, the relativistic Coulomb exchange hole is normalized to $-1$
\begin{eqnarray}
 \int n_{\text{x}}^{\text{C}}(r_{12}) \d\b{r}_{12} = -1,
\end{eqnarray}
but the Breit exchange hole is not. When the density increases from the low-density limit to the high-density limit, its integral varies from $0$ to $1$
\begin{eqnarray}
0 \leq	\int   n_{\text{x}}^{\text{B}}(r_{12})  \d\b{r}_{12} \leq 1,
\end{eqnarray}
which confirms that it should be considered as an effective exchange hole with only purpose to give the Breit exchange energy.

In the non-relativistic limit, corresponding to ${\lambda} =0 $, the Coulomb exchange hole reduces to the well-known non-relativistic form and the Breit exchange hole vanishes
\begin{eqnarray} 
        n_{\text{x}}^{\text{C,NR}}(r_{12}) = -\frac{9}{2} n \frac{j_{1}(k_{\text{F}} r_{12})^{2}}{(k_{\text{F}} r_{12})^{2}}
\text{~~and~~}
 n_{\text{x}}^{\text{B,NR}}(r_{12}) = 0.
\end{eqnarray}
In the opposite ultra-relativistic limit, corresponding to ${\lambda} = 1$, we have 
\begin{eqnarray}
        n_{\text{x}}^{\text{C,UR}}(r_{12}) &=& - \frac{9}{4} n\frac{1}{(k_{\text{F}} r_{12})^{2}}\Bigg[j_{1}(k_{\text{F}} r_{12})^{2} + B_{1}(k_{\text{F}} r_{12})^{2}\Bigg],
\nonumber\\
\text{and}
\nonumber\\
n_{\text{x}}^{\text{B,UR}}(r_{12}) &=&  \frac{9}{2} n\frac{1}{(k_{\text{F}}r_{12})^{2}}j_{1}(k_{\text{F}} r_{12})^{2},
\end{eqnarray}
where we observe again the ultra-relativistic limit of the Breit term is the exact opposite of the non-relativistic limit of the Coulomb contribution.

\begin{figure}[t]
	\centering
	\includegraphics[width=6cm,angle=270]{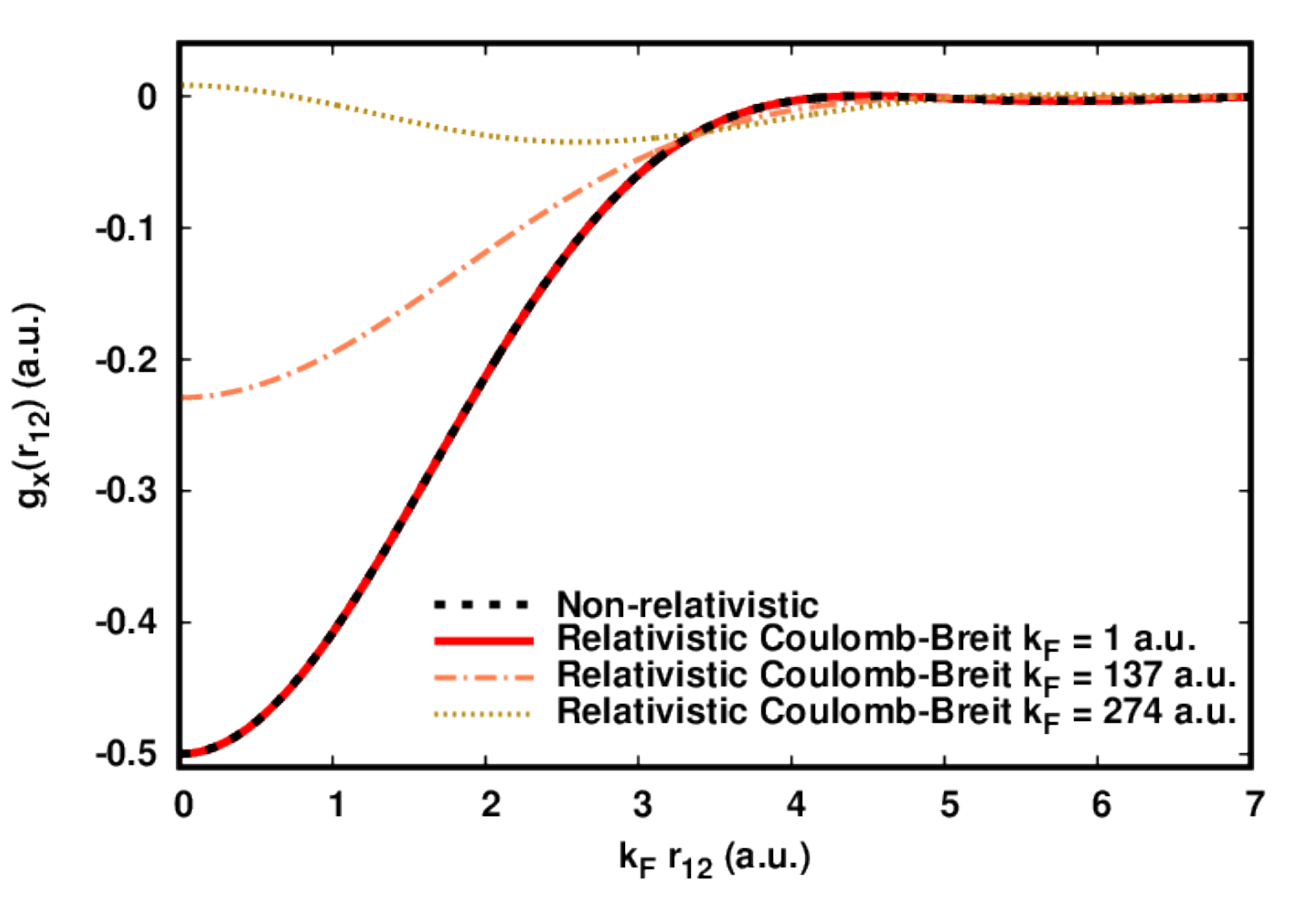}
	\caption{Effective Coulomb-Breit exchange pair-distribution function at different values of $k_\text{F}$ as a function of $k_\text{F} r_{12}$.
}
	\label{fig:pair1}
\end{figure}

In Figure \ref{fig:pair1}, we plot the effective Coulomb-Breit exchange pair-distribution function $g_\text{x}^{\text{CB}}(r_{12}) =  n_{\text{x}}^{\text{CB}}(r_{12})/n$ for different values of the density (or equivalently $k_\text{F}$). For a low value of the density, $k_\text{F}=1$ a.u., $g_\text{x}^{\text{CB}}(r_{12})$ has the usual shape of the exchange hole and is indistinguishable from the non-relativistic calculation, with the on-top value being $g_\text{x}^{\text{CB}}(0)\approx-1/2$. For the higher values of the density, $k_\text{F}=137$ and $274$ u.a., we observe that the Coulomb-Breit exchange hole becomes shallower. In particular, for $k_\text{F}=274$ a.u., the Coulomb-Breit exchange hole is almost flat, with a very small minimum not located at $r_{12}=0$ anymore. These dramatic modifications of the exchange effects due to relativity underline the importance of considering relativistic effects when calculating short-range exchange energies in systems containing high-density regions.

\subsection{Short-range Coulomb-Breit exchange energy per particle}
\label{sec:srrheg}

Since we have previously calculated the Coulomb-Breit exchange hole, the most straightforward way of calculating the short-range Coulomb-Breit exchange energy per particle is to integrate the exchange hole with the short-range electron-electron interaction
\begin{eqnarray}
	{\epsilon}_{\text{x}}^{\text{CB,sr},{\mu}} &=& 
	 \frac{1}{2} \int n_{\text{x}}^{\text{CB}}(r_{12}) ~ w_{\text{ee}}^{\sr,{\mu}}(r_{12}) ~ \d\b{r}_{12}.
\label{epsxCBsrhole}
\end{eqnarray}
Unfortunately, we did not manage to calculate this integral analytically. We thus instead follow the same route as for the calculation of the full-range exchange energy (see Appendix~\ref{app:rheg}), i.e. integrating over space variables first. Similarly to Eq.~(\ref{epsxC}), we obtain the short-range Coulomb exchange energy per particle as a Fourier-space integral where each wave vector spans the volume of the Fermi sphere $V_{k_\text{F}}$
\begin{widetext}
\begin{eqnarray}
{\epsilon}_{\text{x}}^{\text{C,sr},{\mu}} &=& 	-\frac{1}{2n(2{\pi})^{6}} \iint_{V_{k_{\text{F}}}}  ~ \tilde{w}_{\text{ee}}^{\text{sr},{\mu}} (k_{12}) ~ \frac{E_{k_{1}}E_{k_{2}}+(\b{k}_{1}\cdot\b{k}_{2})c^{2}+c^{4}}{E_{k_{1}}E_{k_{2}}} \d\b{k}_{1} \d\b{k}_{2} 
\nonumber\\
&=& \frac{3 k_{\text{F}}}{4{\pi}} \int_{0}^{1} \int_{0}^{1} \d\tilde{k}_{1} \d\tilde{k}_{2} \; \tilde{k}_{1} \tilde{k}_{2} \Bigg(\frac{1}{\sqrt{\tilde{c}^{2} + \tilde{k}_{1}^{2}}\sqrt{\tilde{c}^{2} + \tilde{k}_{2}^{2}}}\Big[\tilde{k}_{1}\tilde{k}_{2} + \Big(e^{-\big(\frac{\tilde{k}_{1}+\tilde{k}_{2}}{2\tilde{\mu}}\big)^{2}} - e^{-\big(\frac{\tilde{k}_{1}-\tilde{k}_{2}}{2\tilde{\mu}}\big)^{2}}\Big)\tilde{\mu}^{2} \Big] 
\nonumber\\ 
&& + \frac{2\tilde{c}^{2} + \tilde{k}_{1}^{2} + \tilde{k}_{2}^{2} +2\sqrt{\tilde{c}^{2} + \tilde{k}_{1}^{2}}\sqrt{\tilde{c}^{2} + \tilde{k}_{2}^{2}} }{4 \sqrt{\tilde{c}^{2} + \tilde{k}_{1}^{2}} \sqrt{\tilde{c}^{2} + \tilde{k}_{2}^{2}}} \left[\text{Ei}\left(-\left(\frac{\tilde{k}_{1}+\tilde{k}_{2}}{2\tilde{\mu}}\right)^{2}\right) - \text{Ei}\left(-\left(\frac{\tilde{k}_{1}-\tilde{k}_{2}}{2\tilde{\mu}}\right)^{2} \right) + \text{ln}\left((\tilde{k}_{1}-\tilde{k}_{2})^{2} \right) - \text{ln}\left((\tilde{k}_{1}+\tilde{k}_{2})^{2} \right) \right] \Bigg), ~~~~~~~
\label{epsCsrintk}
\end{eqnarray}
\end{widetext}
where $\tilde{w}_{\text{ee}}^{\text{sr},{\mu}} (k_{12}) = (4\pi/k_{12}^{2}) (1-e^{-k_{12}^{2}/4{\mu}^{2}})$ is the Fourier transform of the short-range interaction in terms of the relative wave vector $k_{12} =|\b{k}_{12}|$ with $\b{k}_{12} =\b{k}_{1} - \b{k}_{2}$, $\text{Ei}(x) = -\int_{-x}^{\infty} e^{-t}/t ~ \d t$ is the exponential integral function, and we have introduced the scaled variables  $\tilde{\mu} = {\mu}/k_{\text{F}}$, $\tilde{k}_{1} = k_1/k_{\text{F}}$, and $\tilde{k}_{2} = k_2/k_{\text{F}}$. Unfortunately we were unable to straightforwardly calculate the double integral in Eq.~(\ref{epsCsrintk}). To circumvent this difficulty we first make an asymptotic expansion of the integrand for $\tilde{c} \rightarrow {\infty}$ and then integrate term by term to obtain the asymptotic series 
\begin{eqnarray}
{\epsilon}_{\text{x}}^{\text{C,sr},{\mu}} \sim k_\text{F} \sum_{i=0}^{\infty} \frac{{\alpha}_{2i}(\tilde{\mu})}{\tilde{c}^{2i}},
\label{epsCsrasymp}
\end{eqnarray}
where the coefficients ${\alpha}_{2i}(\tilde{\mu})$ can be obtained analytically. Their derivation being quite lengthy, it is discussed in detail in the Supplementary Information~\cite{PaqTou-JJJ-XX-note}. We give here the first three coefficients.
\begin{widetext}
\begin{eqnarray}
{\alpha}_{0}(\tilde{\mu}) &=& \frac{1}{4{\pi}}\Bigg(-3 + 4\sqrt{\pi}~\text{erf}\Big(\frac{1}{\tilde{\mu}}\Big)~\tilde{\mu} + 2\Big(2 e^{-\frac{1}{\tilde{\mu}^{2}}} - 3\Big)~\tilde{\mu}^{2}  -2\Big(e^{-\frac{1}{\tilde{\mu}^{2}}} - 1\Big)~\tilde{\mu}^{4}  \Bigg),
\nonumber\\
{\alpha}_{2}(\tilde{\mu}) &=& \frac{1}{12{\pi}}\Bigg(1 - 6\sqrt{\pi}~\text{erf}\Big(\frac{1}{\tilde{\mu}}\Big)~\tilde{\mu}^{3} - 6\Big(e^{-\frac{1}{\tilde{\mu}^{2}}} - 3\Big)~\tilde{\mu}^{4}  + 12\Big(e^{-\frac{1}{\tilde{\mu}^{2}}} - 1\Big)~\tilde{\mu}^{6}  \Bigg),
\nonumber\\
{\alpha}_{4}(\tilde{\mu}) &=& \frac{1}{240{\pi}}\Bigg(-13 + 12\sqrt{\pi}~\text{erf}\Big(\frac{1}{\tilde{\mu}}\Big)~\Big(8+45\tilde{\mu}^{2}\Big)\tilde{\mu}^{3} + 12\Big(4 e^{-\frac{1}{\tilde{\mu}^{2}}} - 45\Big)~\tilde{\mu}^{4} + 24\Big(13 e^{-\frac{1}{\tilde{\mu}^{2}}} - 40\Big)~\tilde{\mu}^{6} \Big) - 648\Big( e^{-\frac{1}{\tilde{\mu}^{2}}} - 1\Big)~\tilde{\mu}^{8}  \Bigg). 
\end{eqnarray}
\end{widetext}
However, the asymptotic series of Eq.~(\ref{epsCsrasymp}) diverges for $\tilde{c} < 1$, i.e. for $k_{\text{F}} \gtrsim 137$ a.u.. To avoid this divergence, we use a diagonal Pad\'{e} approximant~\cite{BenOrs-BOOK-99} of even order $M$
\begin{eqnarray}
{\epsilon}_{\text{x},\text{Pad\'e}}^{\text{C,sr},{\mu}} = k_\text{F} \frac{\sum_{i=0}^{M/2} A_{2i}(\tilde{\mu})/\tilde{c}^{2i}}{\sum_{i=0}^{M/2} B_{2i}(\tilde{\mu})/\tilde{c}^{2i}},
\label{PadeC}
\end{eqnarray}
with the choice $B_{0}(\tilde{\mu}) = 1$ without loss of generality, and the other coefficients $A_{2i}(\tilde{\mu})$ and $B_{2i}(\tilde{\mu})$ are uniquely determined, for a given $M$, from the coefficients ${\alpha}_{2i}(\tilde{\mu})$ so that the asymptotic expansion of the Pad\'{e} approximant matches the asymptotic expansion of Eq.~(\ref{epsCsrasymp}) up to order $2M$ (see Appendix~\ref{app:pade}). 
For example, we give here the coefficients for the diagonal Pad\'{e} approximant of order 2
\begin{eqnarray}
A_{0}(\tilde{\mu}) &=& {\alpha}_{0}(\tilde{\mu}) ~~~~\text{and}~~~~ A_{2}(\tilde{\mu}) = {\alpha}_{2}(\tilde{\mu}) -\frac{{\alpha}_{0}(\tilde{\mu}){\alpha}_{4}(\tilde{\mu})}{{\alpha}_{2}(\tilde{\mu})}, 
\nonumber\\
&&~~~~~~~~~~~~~~~~~~~~~~                             B_{2}(\tilde{\mu}) = -\frac{{\alpha}_{4}(\tilde{\mu})}{{\alpha}_{2}(\tilde{\mu})}.
\end{eqnarray}

\begin{figure*}[t]
        \centering
        \includegraphics[width=6cm,angle=270]{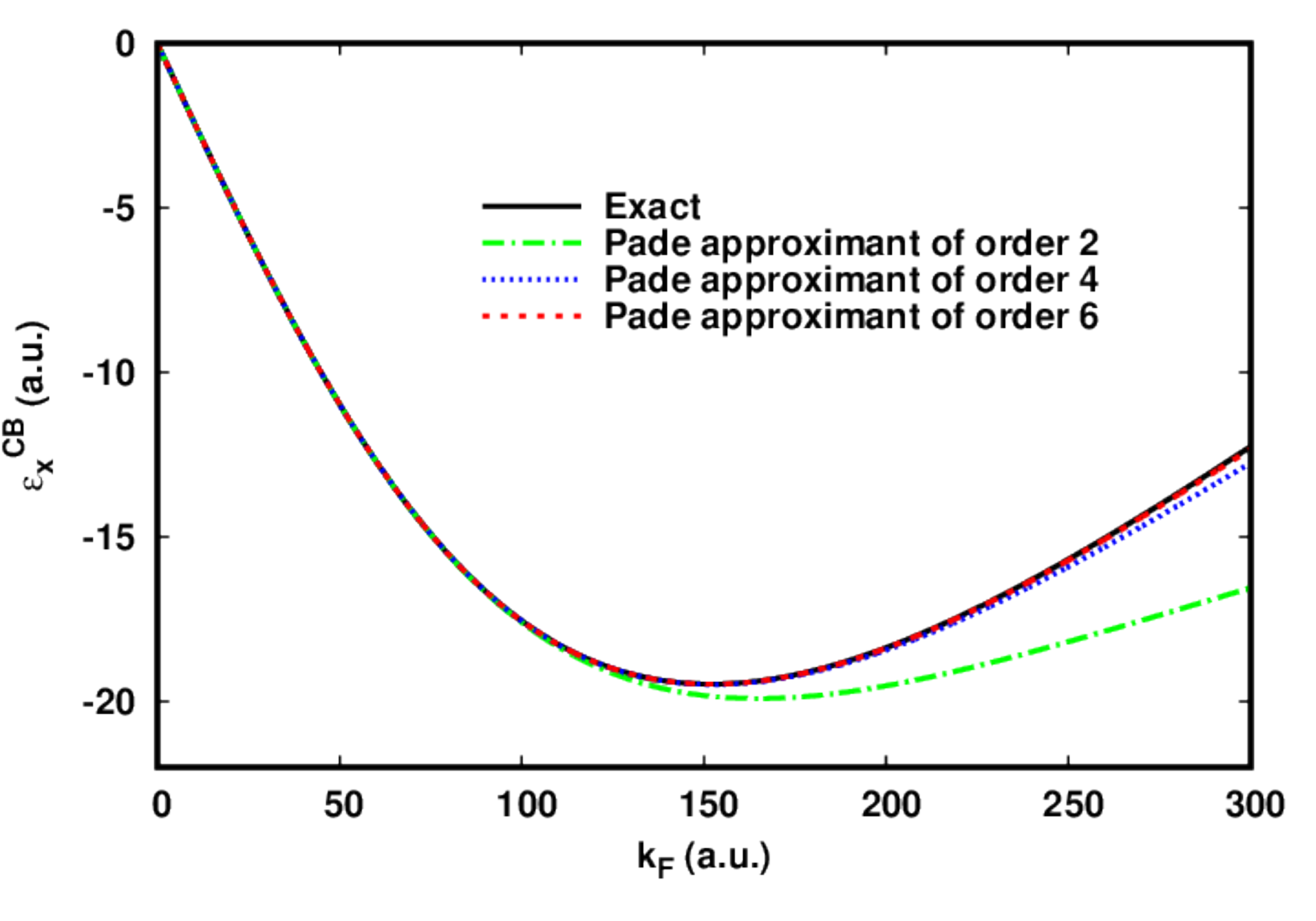}
        \includegraphics[width=6cm,angle=270]{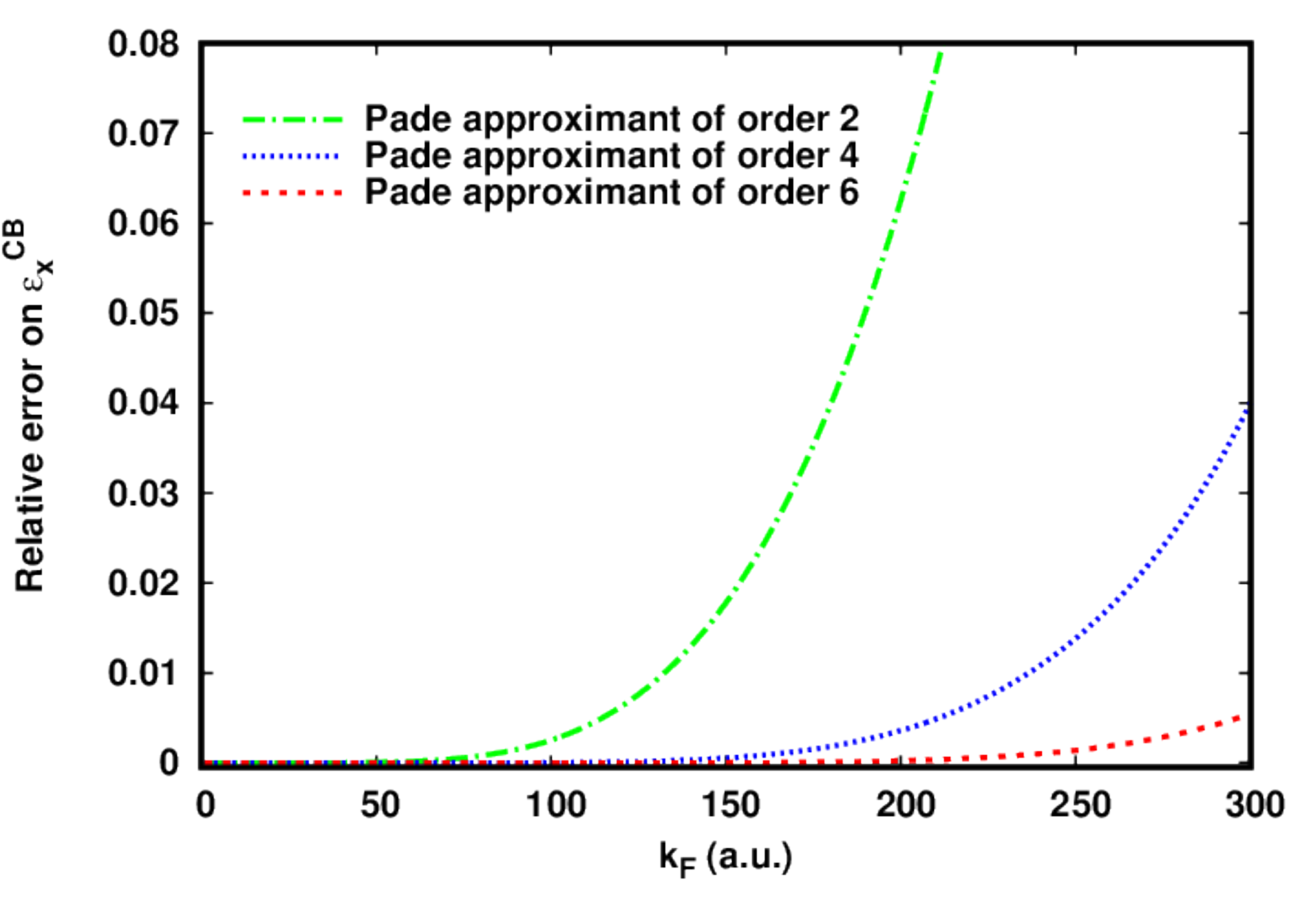}
        \caption{Left plot: Exact full-range Coulomb-Breit exchange energy per particle and its Pad\'{e} approximants of orders 2, 4, and 6 as functions of the Fermi wave vector $k_\text{F}$. Right plot: Relative errors, $\Delta {\epsilon}_{\text{x}}^{\text{CB}} = \left|({\epsilon}_{\text{x},\text{Pad\'e}}^{\text{CB}} - {\epsilon}_{\text{x}}^{\text{CB}})/{\epsilon}_{\text{x}}^{\text{CB}} \right|$,  of the Pad\'{e} approximants compared to the exact full-range Coulomb-Breit exchange energy per particle.}
        \label{fig:Pade}
\end{figure*}

We proceed similarly for the short-range Breit exchange energy per particle. Adapting Eq.~(\ref{epsxB}) to the case of the short-range interaction gives the Fourier-space integral
\begin{widetext}
\begin{eqnarray}
{\epsilon}_{\text{x}}^{\text{B,sr},{\mu}} &=& \frac{1}{2n(2{\pi})^{6}} \iint_{V_{k_{\text{F}}}} ~ \tilde{w}_{\text{ee}}^{\text{sr},{\mu}}(k_{12}) \frac{c^{2}}{E_{k_{1}}E_{k_{2}}} \Bigg( \frac{E_{k_{2}}+c^{2}}{E_{k_{1}}+c^{2}}k_{1}^{2} + \frac{E_{k_{1}}+c^{2}}{E_{k_{2}}+c^{2}} k_{2}^{2} 
\Bigg)\d\b{k}_{1} \d\b{k}_{2} 
\nonumber\\
&=& \frac{3 k_{\text{F}}}{4{\pi}} \int_{0}^{1} \! \int_{0}^{1} \! \d\tilde{k}_{1} \d\tilde{k}_{2} \; \tilde{k}_{1} \tilde{k}_{2} \frac{\tilde{c}^{2}  - \sqrt{\tilde{c}^{2} + \tilde{k}_{1}^{2}}\sqrt{\tilde{c}^{2} + \tilde{k}_{2}^{2}} }{\sqrt{\tilde{c}^{2} + \tilde{k}_{1}^{2}} \sqrt{\tilde{c}^{2} + \tilde{k}_{2}^{2}}} \left[\text{Ei}\left(-\left(\frac{\tilde{k}_{1}+\tilde{k}_{2}}{2\tilde{\mu}}\right)^{2}\right) - \text{Ei}\left(-\left(\frac{\tilde{k}_{1}-\tilde{k}_{2}}{2\tilde{\mu}}\right)^{2} \right) + \text{ln}\left((\tilde{k}_{1}-\tilde{k}_{2})^{2} \right) - \text{ln}\left((\tilde{k}_{1}+\tilde{k}_{2})^{2} \right) \right], ~~~~~~
\label{epsBsrintk}
\end{eqnarray}
	\end{widetext}
which, after asymptotically expanding the integrand for $\tilde{c} \rightarrow {\infty}$ and integrating term by term, turns into the asymptotic series
\begin{eqnarray}
	{\epsilon}_{\text{x}}^{\text{B,sr},\mu} \sim k_\text{F} \sum_{i=0}^{\infty} \frac{{\beta}_{2i}(\tilde{{\mu}})}{\tilde{c}^{2i}},
\label{epsBsrasymp}
\end{eqnarray}
where the coefficients ${\beta}_{2i}(\tilde{\mu})$ can be obtained analytically. The first three of them are
\begin{widetext}
\begin{eqnarray}
{\beta}_{0}(\tilde{{\mu}}) &=& 0,
\nonumber\\
{\beta}_{2}(\tilde{{\mu}}) &=& \frac{1}{60{\pi}} \Bigg(25 -12\sqrt{\pi}\Big(3+5\tilde{\mu}^{2}\Big)~\text{erf}\Big(\frac{1}{\tilde{\mu}}\Big)~\tilde{\mu} + 18\Big(5 - 2e^{-\frac{1}{\tilde{\mu}^{2}}}\Big)\tilde{\mu}^{2} + 6\Big(15 - 7 e^{-\frac{1}{\tilde{\mu}^{2}}}\Big)\tilde{\mu}^{4}  -48\Big(1 - e^{-\frac{1}{\tilde{\mu}^{2}}}\Big)\tilde{\mu}^{6}
\Bigg),
\nonumber\\
	{\beta}_{4}(\tilde{{\mu}}) &=& \frac{1}{280{\pi}} \Bigg(-77 +12\sqrt{\pi}\Big(10+42\tilde{\mu}^{2}+105\tilde{\mu}^{4}\Big)~\text{erf}\Big(\frac{1}{\tilde{\mu}}\Big)~\tilde{\mu} - 60\Big(7 - 2e^{-\frac{1}{\tilde{\mu}^{2}}}\Big)\tilde{\mu}^{2} - 12\Big(140 - 37 e^{-\frac{1}{\tilde{\mu}^{2}}}\Big)\tilde{\mu}^{4}  
\nonumber\\
&& ~~~~~~~~~~~~~~~~~~~~~~~~~~~~~~~~~~~~~~~~~~~~~~~~~~~~~~~~~~~~~~~~~~~~~~~~~~~~~~~~~~~~~~~ -24\Big(70 - 37 e^{-\frac{1}{\tilde{\mu}^{2}}}\Big)\tilde{\mu}^{6} +792\Big(1 - e^{-\frac{1}{\tilde{\mu}^{2}}}\Big)\tilde{\mu}^{8} \Bigg). 
\end{eqnarray}
\end{widetext}
The fact that ${\beta}_{0}(\tilde{{\mu}}) =  0$ corresponds to the fact that the Breit exchange energy vanishes in the non-relativistic limit. Again, the asymptotic series of Eq.~(\ref{epsBsrasymp}) diverges for $\tilde{c} < 1$, so we construct a diagonal Pad\'{e} approximant of even order $M$
\begin{eqnarray}
{\epsilon}_{\text{x},\text{Pad\'e}}^{\text{B,sr},{\mu}} = k_\text{F} \frac{\sum_{i=0}^{M/2} C_{2i}(\tilde{\mu})/\tilde{c}^{2i}}{\sum_{i=0}^{M/2} D_{2i}(\tilde{\mu})/\tilde{c}^{2i}},
\label{PadeB}
\end{eqnarray}
with $D_{0}(\tilde{\mu}) = 1$ and the other coefficients $C_{2i}(\tilde{\mu})$ and $D_{2i}(\tilde{\mu})$ are uniquely determined, for a given $M$, from the coefficients ${\beta}_{2i}(\tilde{\mu})$.
We give here the coefficients of the diagonal Pad\'{e} approximant of order 2
\begin{eqnarray}
C_{0}(\tilde{\mu}) &=& {\beta}_{0}(\tilde{\mu}) ~~~~\text{and}~~~~ C_{2}(\tilde{\mu}) = {\beta}_{2}(\tilde{\mu}) -\frac{{\beta}_{0}(\tilde{\mu}){\beta}_{4}(\tilde{\mu})}{{\beta}_{2}(\tilde{\mu})},
\nonumber\\
&&~~~~~~~~~~~~~~~~~~~~~~ D_{2}(\tilde{\mu}) = -\frac{{\beta}_{4}(\tilde{\mu})}{{\beta}_{2}(\tilde{\mu})}.
\end{eqnarray}
The complete expressions of the large-$\ct$ expansions in Eqs.~(\ref{epsCsrasymp}) and~(\ref{epsBsrasymp}) and their associated Pad\'e approximants in Eqs.~(\ref{PadeC}) and~(\ref{PadeB}) are explicitly given up to an arbitrary order in the Mathematica notebook available in the Supplementary Information~\cite{PaqTou-JJJ-XX-note}. 

We want to check the accuracy of these Pad\'{e} approximants to the short-range Coulomb-Breit exchange energy per particle and at which order $M$ we can truncate the expansion. For this, we may check for the most demanding case of $\mu=0$ corresponding to the full-range interaction and for which we know the exact exchange energy per particle [Eqs.~(\ref{epsxCtext}) and~(\ref{epsxBtext})]. We thus plot in Figure \ref{fig:Pade} the exact full-range Coulomb-Breit exchange energy per particle and its Pad\'{e} approximants of orders 2, 4, and 6, as well as their relative errors $\Delta {\epsilon}_{\text{x}}^{\text{CB}} = \left|({\epsilon}_{\text{x},\text{Pad\'e}}^{\text{CB}} - {\epsilon}_{\text{x}}^{\text{CB}})/{\epsilon}_{\text{x}}^{\text{CB}} \right|$, as functions of the Fermi wave vector $k_\text{F}$. We note in passing that the Pad\'e approximants can be calculated either for the Coulomb and Breit exchange energies per particle separately, or directly for the total Coulomb-Breit exchange energy per particle, as done here for the plot. It turns out that these two ways of proceeding give very similar Pad\'e approximants for the Coulomb-Breit exchange energy per particle, for example differing by at most about $10^{-4}$ \% for the Pad\'e approximants of order 6.
As seen in Figure \ref{fig:Pade}, the error of all Pad\'{e} approximants naturally increases with the density. The error of the Pad\'{e} approximant of order 2 increases most rapidly with the density, the relative error exceeding 5\% as soon as $k_{\text{F}} \gtrsim 200$ a.u.. The relative error in the Pad\'{e} approximant of order 4 is less than 0.5\% until about $k_{\text{F}} = 200$ a.u. and increases to about 4\% at $k_{\text{F}} = 300$ a.u.. The Pad\'{e} approximant of order 6 has a maximal relative error of 0.5\% at $k_{\text{F}} = 300$ a.u.. We thus conclude that the Pad\'{e} approximant of order 6 provides an excellent approximation for all the density range in which we are interested. We have also explicitly checked that this accuracy of the Pad\'{e} approximant of order 6 is preserved for the short-range Coulomb-Breit exchange energy per particle for a non-zero value of $\mu$ by comparing to results obtained by numerical integration in Eq.~(\ref{epsxCBsrhole}). Hence, numerically, the Pad\'{e} approximants appear to rapidly converge to the correct limit as $M\to \infty$.

\begin{figure}[t]
        \centering
        \includegraphics[width=5cm,angle=270]{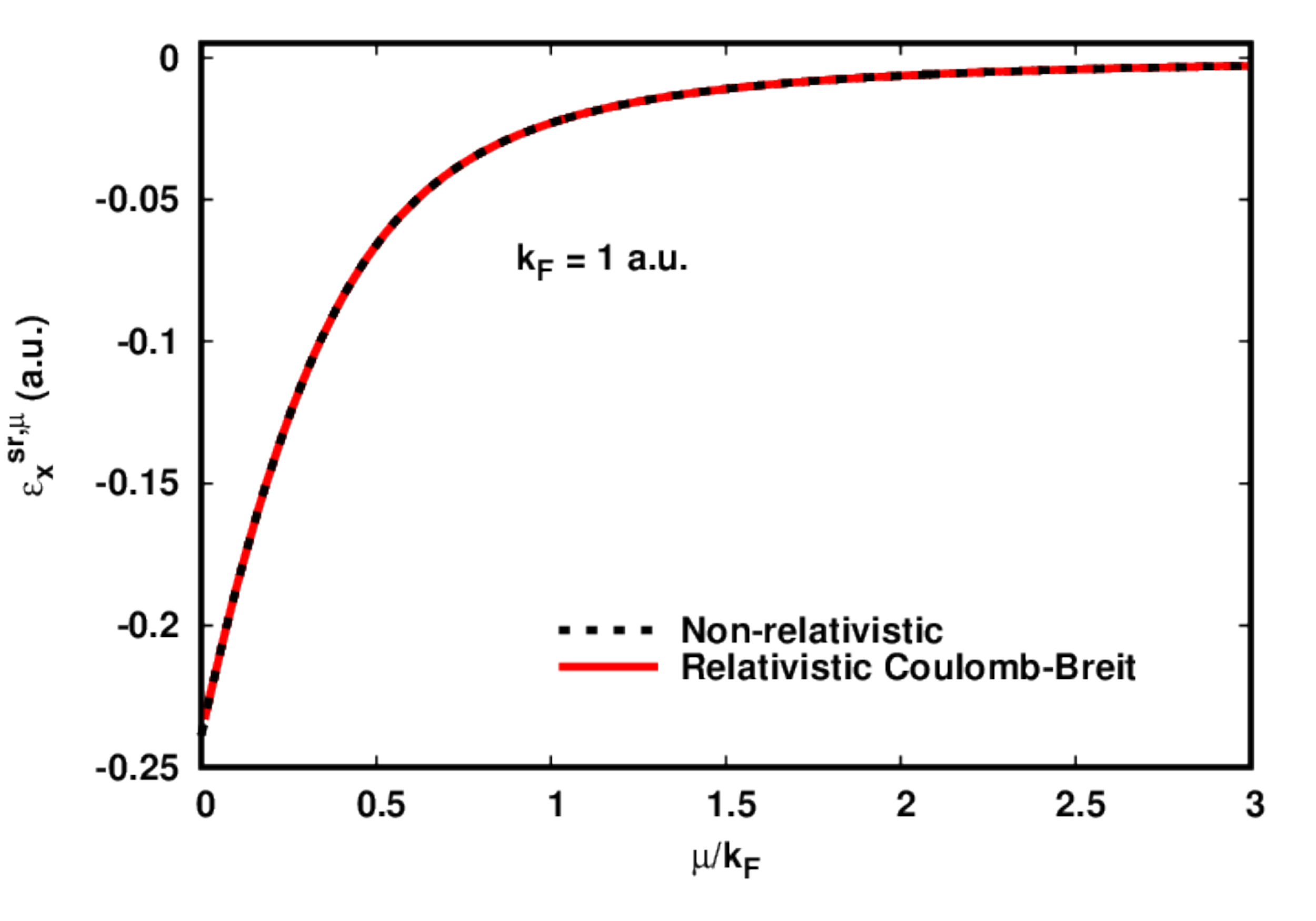}
        \includegraphics[width=5cm,angle=270]{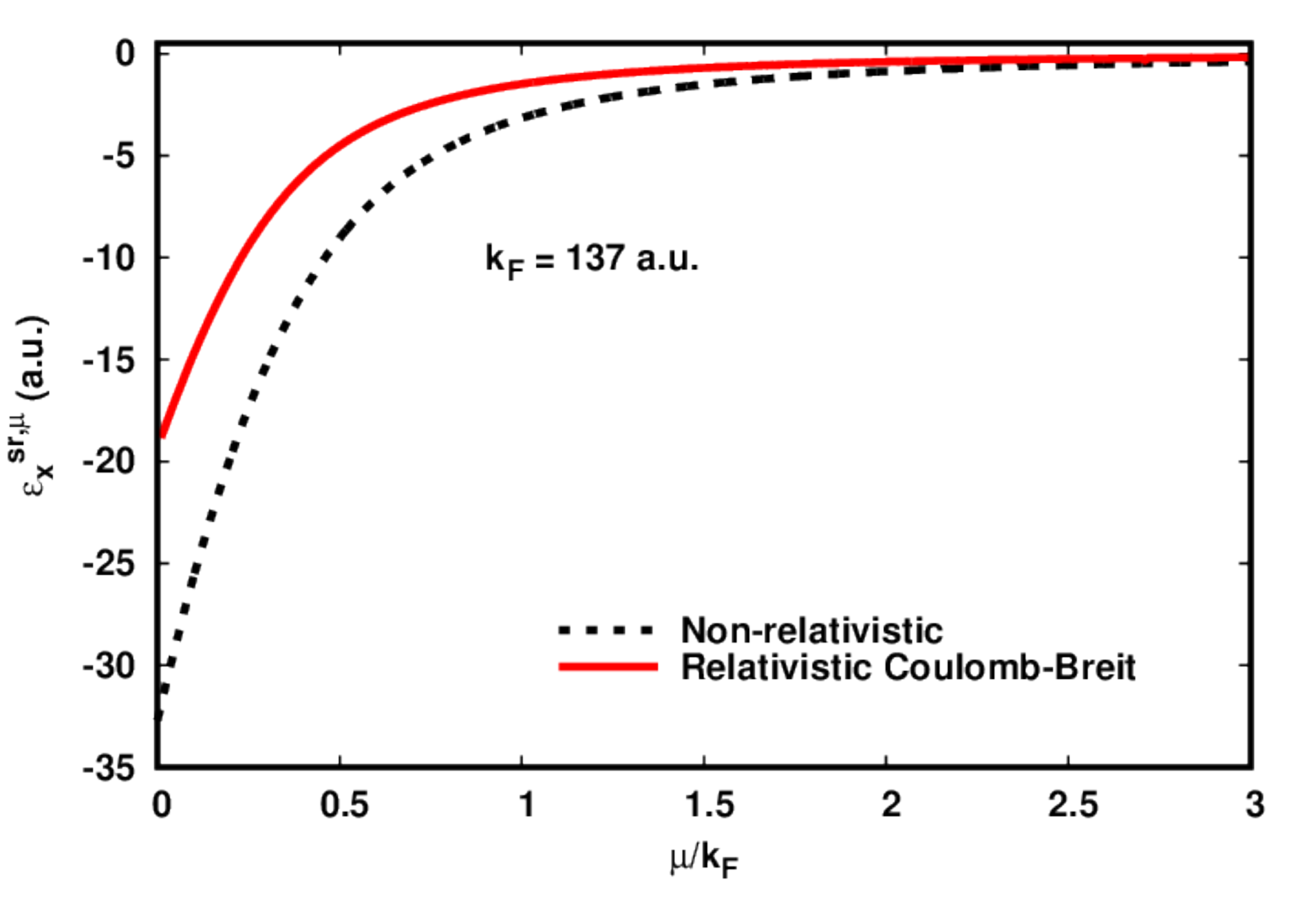}
        \includegraphics[width=5cm,angle=270]{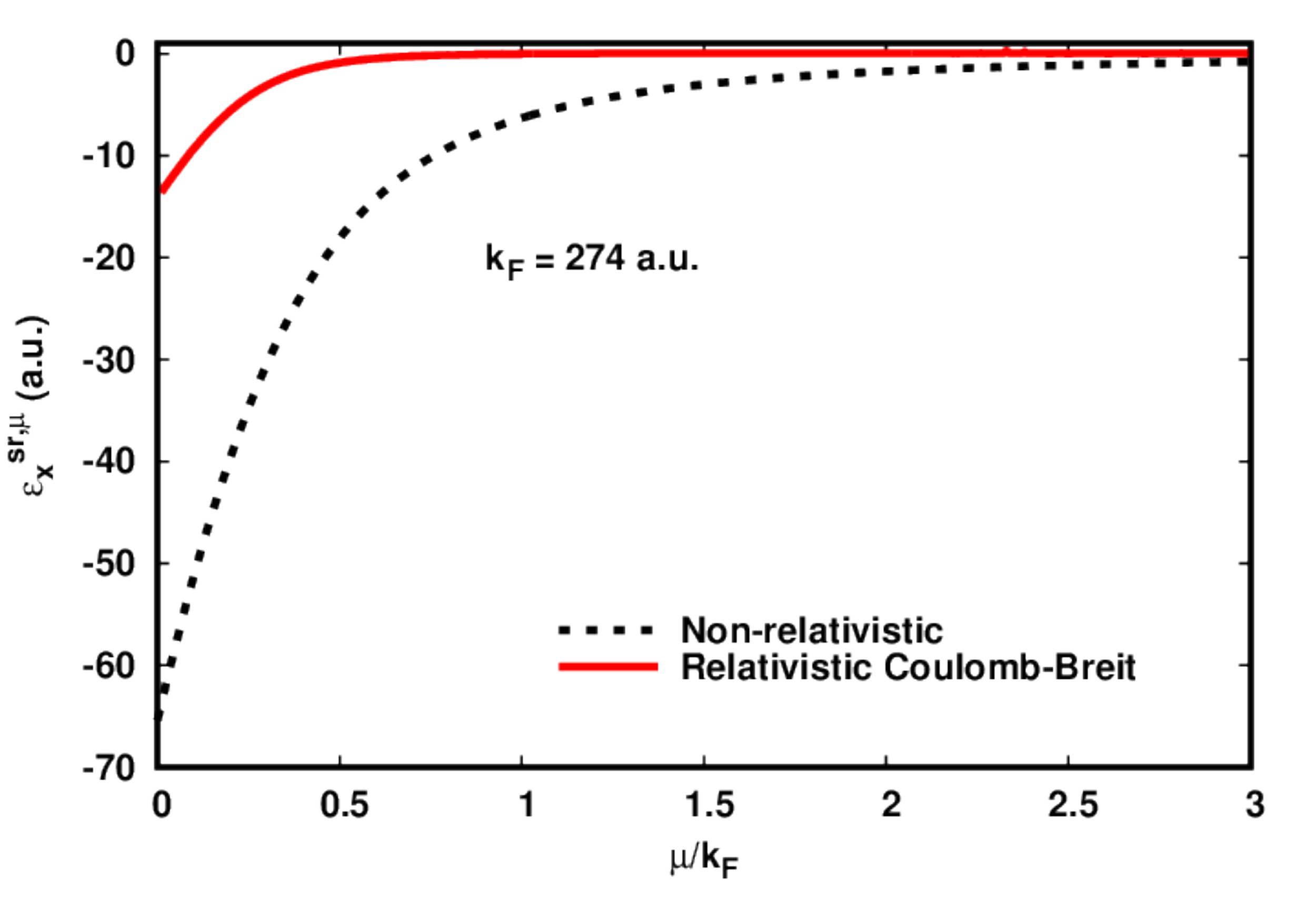}
	\caption{Non-relativistic and relativistic Coulomb-Breit short-range exchange energies per particle as functions of $\mu/k_\text{F}$. The relativistic Coulomb-Breit short-range exchange energy is obtained from the Pad\'e approximant of order 6.}
        \label{fig:ShortPade}
\end{figure}

In Figure \ref{fig:ShortPade} we plot the relativistic short-range Coulomb-Breit exchange energy per particle, using the Pad\'{e} approximant of order 6, as a function of the scaled range-separation parameter $\mu/k_\text{F}$ for different values of $k_\text{F}$, and compare it to the non-relativistic short-range exchange energy per particle (whose expression can be found in Refs.~\onlinecite{Sav-INC-96,TouSavFla-IJQC-04}). For $k_\text{F}=1$ a.u., the relativistic exchange energy per particle is indistinguishable from its non-relativistic counterpart, for all values of $\mu$. For higher values of the Fermi wave vector, $k_{\text{F}}=137$ and $274$ a.u., the relativistic short-range Coulomb-Breit exchange energy per particle becomes much smaller, in absolute value, than the non-relativistic short-range exchange energy per particle. Also, it appears that, for large $k_{\text{F}}$, the relativistic short-range exchange energy per particle goes to zero when increasing ${\mu}/k_{\text{F}}$ significantly faster than does its non-relativistic analogue.

\subsection{Small- and large-$\mu$ expansions and simple approximation for the short-range Coulomb-Breit exchange energy per particle}
\label{sec:expansions}

We have determined the short-range Coulomb and Breit exchange energies per particle ${\epsilon}_{\text{x}}^{\text{C,sr},{\mu}}$ and ${\epsilon}_{\text{x}}^{\text{B,sr},{\mu}}$ in the form of Pad\'e approximants with quite complicated coefficients. We now derive simple expressions valid for small and large values of the range-separation parameter ${\mu}$ and use them to construct simple approximations for ${\epsilon}_{\text{x}}^{\text{C,sr},{\mu}}$ and ${\epsilon}_{\text{x}}^{\text{B,sr},{\mu}}$.

In order to obtain the expansions of ${\epsilon}_{\text{x}}^{\text{C,sr},{\mu}}$ and ${\epsilon}_{\text{x}}^{\text{B,sr},{\mu}}$ for ${\mu}\to 0$, we start from the asymptotic series in Eqs.~(\ref{epsCsrasymp}) and~(\ref{epsBsrasymp}), expand them with respect to $\mu$, and extract the ${\mu}$ and ${\mu}^{2}$ terms. 
For the short-range Coulomb exchange energy per particle, we obtain directly expressions independent of ${\ct}$ for the linear and quadratic terms in ${\mu}$
\begin{eqnarray}
	{\epsilon}_{\text{x}}^{\text{C,sr},{\mu}} &=& {\epsilon}_{\text{x}}^{\text{C}} + \frac{\mu}{\sqrt{\pi}} -\frac{3 ~{\mu}^{2}}{2{\pi} ~k_{\text{F}}} +  O\left({\mu}^{3}\right),
\label{Coulombsmallmu}
\end{eqnarray}
and for the short-range Breit exchange energy per particle we obtain expansions in ${\ct}$ for the linear and quadratic terms in ${\mu}$
\begin{eqnarray}
	{\epsilon}_{\text{x}}^{\text{B,sr},{\mu}} &=& {\epsilon}_{\text{x}}^{\text{B}} - \frac{\mu}{\sqrt{\pi}}\left( \frac{3}{5{\ct}^{2}} - \frac{3}{7{\ct}^{4}} + \frac{3}{9{\ct}^{6}} -\frac{3}{11{\ct}^{8}} + O\left(\frac{1}{{\ct}^{10}}\right)\right) 
\nonumber\\
&+& \frac{3 ~{\mu}^{2}}{2{\pi} ~k_{\text{F}}} \left( \frac{1}{{\ct}^{2}} - \frac{1}{{\ct}^{4}} + \frac{1}{{\ct}^{6}} -\frac{1}{{\ct}^{8}} + O\left(\frac{1}{{\ct}^{10}}\right) \right) + O\left({\mu}^{3}\right). \;\;\;\;\;
\label{epsxBsrexpandmu}
\end{eqnarray}
In this expression, the series in $1/\ct$ can be exactly summed, and we can write Eq.~(\ref{epsxBsrexpandmu}) as
\begin{eqnarray}
 	{\epsilon}_{\text{x}}^{\text{B,sr},{\mu}} &=& {\epsilon}_{\text{x}}^{\text{B}} - \frac{\mu}{\sqrt{\pi}}\left(1 - f(\ct) \right) + \frac{3 ~{\mu}^{2}}{2{\pi} ~k_{\text{F}}} \left(1 - g(\ct) \right) + O\left({\mu}^{3}\right),
\nonumber\\
\label{Breitsmallmu}
\end{eqnarray}
where
\begin{eqnarray}
f(\ct) =	3~{\ct}^{2} - 3~{\ct}^{3}~\text{arctan}\Big(\frac{1}{\ct}\Big),
\end{eqnarray}
and
\begin{eqnarray}
g(\ct) =	\frac{1}{1+\frac{1}{{\ct}^{2}}}.
\end{eqnarray}
We note that the linear term in $\mu$ in Eq.~(\ref{Breitsmallmu}) can also been found by inserting into Eq.~(\ref{epsBsrintk}) the Fourier transform of the linear term in $\mu$ of the expansion of the short-range interaction. However, such an approach cannot be used for higher-order terms in $\mu$ because these terms lead to divergences when inserting into Eq.~(\ref{epsBsrintk}). The expansion of the short-range Coulomb-Breit exchange energy per particle for ${\mu}\to 0$ is thus
\begin{eqnarray}
	{\epsilon}_{\text{x}}^{\text{CB,sr},\mu} &=& {\epsilon}_{\text{x}}^{\text{CB}} + \frac{\mu}{\sqrt{\pi}}f(\ct) - \frac{3 ~{\mu}^{2}}{2{\pi} ~k_{\text{F}}} g(\ct) + O\left({\mu}^{3}\right).
\label{CoulombBreitsmallmu}
\end{eqnarray}
Therefore, near $\mu=0$, the linear and quadratic terms in $\mu$ of ${\epsilon}_{\text{x}}^{\text{CB,sr},\mu}$ depend on $\ct$. In the non-relativistic limit, we have $f(\ct \to \infty)=1$ and $g(\ct \to \infty)=1$, and thus we correctly recover the non-relativistic expansion $\mu/\sqrt{\pi} - 3 ~{\mu}^{2}/(2{\pi}k_{\text{F}})$~\cite{TouSavFla-IJQC-04,TouColSav-PRA-04}. In the ultra-relativistic limit, we have $f(\ct \to 0)=0$ and $g(\ct \to 0)=0$, and thus the linear and quadratic terms in $\mu$ vanish.

In order to obtain the expansions of ${\epsilon}_{\text{x}}^{\text{C,sr},{\mu}}$ and ${\epsilon}_{\text{x}}^{\text{B,sr},{\mu}}$ for ${\mu}\to\infty$, we start from the asymptotic expansion of the Fourier transform of the short-range interaction
\begin{eqnarray} 
	\tilde{w}_{\text{ee}}^{\sr,{\mu}}(k_{12})  &=& \frac{\pi}{{\mu}^{2}} +O\left(\frac{1}{{\mu}^{4}}\right).
\label{wsrkmuinfty}
\end{eqnarray} 
Inserting then Eq.~(\ref{wsrkmuinfty}) into Eq.~(\ref{epsCsrintk}) leads to the expansion of the short-range Coulomb exchange energy per particle
\begin{eqnarray}
	{\epsilon}_{\text{x}}^{\text{C,sr},{\mu}} &=& -\frac{k_{\text{F}}^3}{24{\pi}{{\mu}}^{2}}\left( 1 + h(\ct) \right) + O\left(\frac{1}{{\mu}^{4}}\right),
\label{Coulomblargemu}
\end{eqnarray}
with
\begin{eqnarray}
	h(\ct) &=& \frac{9}{4}~({\ct}^{2} + {\ct}^{4}) 
\nonumber\\
&&- \frac{9}{4}~{\ct}^{4} \text{arcsinh}\Big(\frac{1}{\ct}\Big) \Bigg(2\sqrt{1+{\ct}^{2}} - {\ct}^{2}~\text{arcsinh}\Big(\frac{1}{\ct}\Big)\Bigg).
\end{eqnarray}
Similarly, inserting Eq.~(\ref{wsrkmuinfty}) into Eq.~(\ref{epsBsrintk}) leads to the expansion of the short-range Breit exchange energy per particle
\begin{eqnarray}
	{\epsilon}_{\text{x}}^{\text{B,sr},{\mu}} &=& \frac{k_{\text{F}}^3}{12{\pi}{{\mu}}^{2}}\left(1 - h(\ct) \right)  + O\left(\frac{1}{{\mu}^{4}}\right).
\label{Breitlargemu}
\end{eqnarray}
The asymptotic expansion of the short-range Coulomb-Breit exchange energy per particle for ${\mu}\to \infty$ is thus
\begin{eqnarray}
	{\epsilon}_{\text{x}}^{\text{CB,sr},{\mu}} = -\frac{k_{\text{F}}^3}{24{\pi}{{\mu}}^{2}}\left( 3 h(\ct) -1\right) + O\left(\frac{1}{{\mu}^{4}}\right).
\label{CoulombBreitlargemu}
\end{eqnarray}
In the non-relativistic limit, we have $h(\ct \to \infty)=1$ and we thus recover the leading term of the asymptotic expansion of the non-relativistic short-range exchange energy per particle, $-k_{\text{F}}^3/(12\pi\mu^2)$~\cite{TouSavFla-IJQC-04,TouColSav-PRA-04}. In the ultra-relativistic limit, we have $h(\ct \to 0)=0$ and thus the leading term is $k_{\text{F}}^3/(24\pi\mu^2)$.

\begin{figure}[t]
\centering
\includegraphics[width=6cm,angle=270]{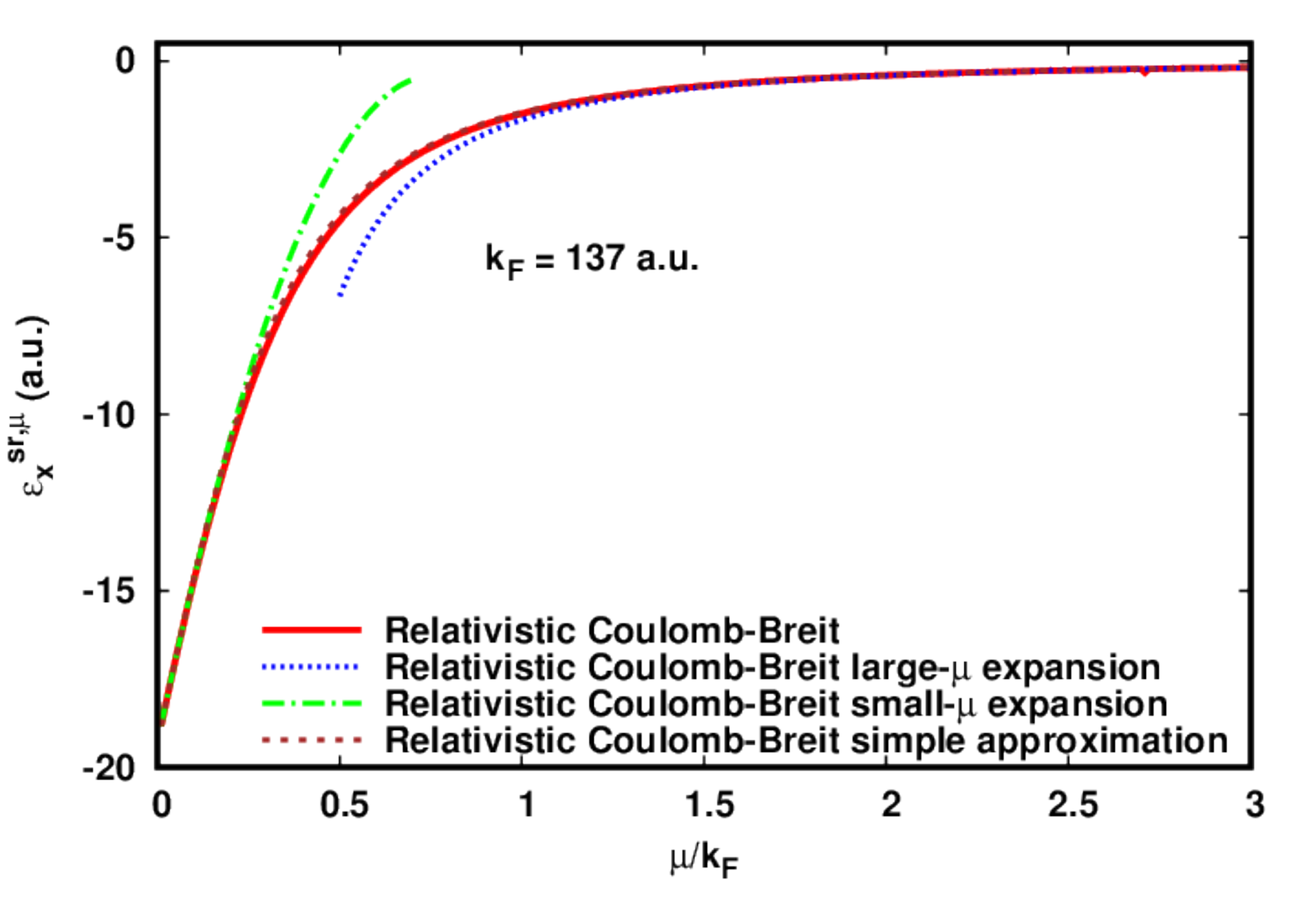}
	\caption{Relativistic Coulomb-Breit short-range exchange energy per particle, its small- and large-${\mu}$ expansions of Eqs.~(\ref{CoulombBreitsmallmu}) and~(\ref{CoulombBreitlargemu}), and its simple approximation of Eqs.~(\ref{simpleapproxC}) and~(\ref{simpleapproxB}) as functions of ${\mu}/k_{\text{F}}$. The reference relativistic Coulomb-Breit short-range exchange energy per particle is obtained from the Pad\'e approximant of order 6.}
\label{fig:Limits}
\end{figure}

We can see on Figure \ref{fig:Limits} that these small- and large-${\mu}$ expansions indeed reproduce well, at the scale of the plot, the short-range Coulomb-Breit exchange energy per particle for $\mu/k_\text{F} \lesssim 0.25$ and $\mu/k_\text{F} \gtrsim 1$, respectively.

We propose now simple approximations for ${\epsilon}_{\text{x}}^{\text{C,sr},{\mu}}$ and ${\epsilon}_{\text{x}}^{\text{B,sr},{\mu}}$ as rational interpolations between these small- and large-${\mu}$ expansions. The short-range Coulomb exchange energy per particle can be approximated by
\begin{eqnarray}
{\epsilon}_{\text{x}}^{\text{C,sr},{\mu}} \approx \frac{{\epsilon}_{\text{x}}^{\text{C}} + d^\text{C}{\mu}}{1 + a^\text{C} {\mu}+ b^\text{C}{\mu}^{2} + c^\text{C}{\mu}^{3}},
\label{simpleapproxC}
\end{eqnarray}
where the coefficients are determined to recover the $\mu$ and $\mu^2$ terms of the small-$\mu$ expansion of Eq.~(\ref{Coulombsmallmu}) and the $1/\mu^2$ term and the vanishing $1/\mu^3$ term of the large-$\mu$ expansion of Eq.~(\ref{Coulomblargemu})
\begin{eqnarray}
a^\text{C} &=& \frac{3}{2{\pi}^{1/2}k_{\text{F}}} + \frac{24{\pi}^{3/2}}{k_{\text{F}}^{3}(1+h(\ct))} ({\epsilon}_{\text{x}}^{\text{C}})^{2},
\nonumber\\
b^\text{C} &=& -\frac{24{\pi}}{k_{\text{F}}^{3}(1+h(\ct))}{\epsilon}_{\text{x}}^{\text{C}},
\nonumber\\
c^\text{C} &=& - \frac{24{\pi}}{k_{\text{F}}^{3}(1+h(\ct))}d^\text{C}, 
\nonumber\\
d^\text{C} &=& \frac{1}{{\pi}^{1/2}} + a^\text{C}{\epsilon}_{\text{x}}^{\text{C}}.
\end{eqnarray}
In a similar way, we propose to approximate the short-range Breit exchange energy per particle by
\begin{eqnarray}
{\epsilon}_{\text{x}}^{\text{B,sr},{\mu}} \approx \frac{{\epsilon}_{\text{x}}^{\text{B}} + d^\text{B}{\mu}}{1 + a^\text{B} {\mu}+ b^\text{B}{\mu}^{2} + c^\text{B}{\mu}^{3}},
\label{simpleapproxB}
\end{eqnarray}
where the coefficients are again determined by imposing the small-$\mu$ expansion of Eq.~(\ref{Breitsmallmu}) and the large-$\mu$ expansion of Eq.~(\ref{Breitlargemu}).
\begin{eqnarray}
a^\text{B} &=& \frac{3(1-g(\ct))}{2{\pi}^{1/2}k_{\text{F}}(1-f(\ct))} + \frac{12{\pi}^{3/2}}{k_{\text{F}}^{3}(1-h(\ct))(1-f(\ct))}({\epsilon}_{\text{x}}^{\text{B}})^{2},
\nonumber\\
b^\text{B} &=& \frac{12{\pi}}{k_{\text{F}}^{3}(1-h(\ct))}{\epsilon}_{\text{x}}^{\text{B}},
\nonumber\\
c^\text{B} &=& \frac{12{\pi}}{k_{\text{F}}^{3}(1-h(\ct))} d^\text{B},
\nonumber\\
d^\text{B} &=& -\frac{(1-f(\ct))}{{\pi}^{1/2}} + a^{B}{\epsilon}_{\text{x}}^{\text{B}}.
\end{eqnarray}

Using the same form of approximation on directly the short-range Coulomb-Breit exchange energy per particle can lead to poles in ${\mu}$, thus it is preferable to separately construct approximations for the Coulomb and Breit contributions and to sum them to get an approximation of the short-range Coulomb-Breit exchange energy per particle. This simple approximation is reported on Figure \ref{fig:Limits} where it can be seen that it is a quite good approximation to the accurate short-range Coulomb-Breit exchange energy per particle given by the Pad\'e approximant of order 6. The relative error of this simple approximation is less than 5\% for all values of $k_{\text{F}}$ and ${\mu}$. For higher accuracy, one could construct rational interpolations with more coefficients determined from higher-order terms in the small- and large-$\mu$ expansions. However, while the large-$\mu$ expansion could easily be obtained at an arbitrary order, higher-order terms in the small-$\mu$ expansion would be obtained as large-$c$ asymptotic series that might not be always easy to sum into a closed form.

\section{Conclusions}
\label{sec:conclusions}
 
In this work, we have considered the extension of RS-DFT to a four-component relativistic framework using a Dirac-Coulomb-Breit Hamiltonian, and we have constructed a short-range LDA exchange density functional based on calculations on the RHEG with a modified electron-electron interaction. More specifically, we have provided the relativistic short-range Coulomb and Breit exchange energies per particle of the RHEG in the form of Pad\'e approximants [Eqs.~(\ref{PadeC}) and~(\ref{PadeB})], constructed from large-$c$ asymptotic expansions (but without involving expansions with respect to $\mu$) and which are systematically improvable to arbitrary accuracy. These quantities, as well as the associated effective Coulomb-Breit exchange hole of the RHEG, show the important impact of relativity on short-range exchange effects for high densities. We have also provided simpler approximations for the relativistic short-range Coulomb and Breit exchange energies per particle of the RHEG in the form of rational interpolations [Eqs.~(\ref{simpleapproxC}) and~(\ref{simpleapproxB})] constructed from the exact small- and large-$\mu$ expansions (but without involving expansions with respect to $c$) of the short-range exchange energies per particle, which can also be used when a limited accuracy is sufficient (relative error less than 5\%).

Possible continuation of this work includes the construction of a relativistic short-range LDA correlation density functional (even though relativistic effects are expected to be much smaller for correlation than for exchange), the construction of relativistic short-range GGA functionals, the inclusion of the dependence on the current for open-shell systems, and the implementation of these functionals in a four-component relativistic RS-DFT program for tests on atomic and molecular systems. In particular, for compounds with heavy elements, including relativistic effects in short-range functionals should have significant effects on quantities sensitive to atomic cores, such as total energies, NMR and EPR parameters, or X-ray spectra.

\appendix
\begin{widetext}
\section{Coulomb-Breit exchange energy per particle and exchange hole of the relativistic homogeneous electron gas}
\label{app:rheg}

In this appendix, we review the calculation of the Coulomb-Breit exchange energy per particle and the exchange hole of the RHEG.

\subsection{Coulomb-Breit exchange energy per particle}

The Coulomb exchange energy per particle of the RHEG can be straightforwardly calculated by summing over spins, integrating over each space coordinates in the volume $V$, and finally integrating over each wave vector in the volume of the Fermi sphere $V_{k_\text{F}}$ 
\begin{eqnarray}
{\epsilon}_{\text{x}}^{\text{C}} &=&  -\frac{1}{2N}\frac{V^{2}}{(2{\pi})^{6}} \iint_{V_{k_{\text{F}}}} \iint_{V}  \sum_{{\sigma}_{1},{\sigma}_{2}={\downarrow},{\uparrow}} w_{\text{ee}}(r_{12}) \; {\psi}_{\b{k}_{1},{\sigma}_{1}}^{\dagger}(\b{r}_1){\psi}_{\b{k}_{2},{\sigma}_{2}}(\b{r}_1){\psi}_{\b{k}_{2},{\sigma}_{2}}^{\dagger}(\b{r}_2){\psi}_{\b{k}_{1},{\sigma}_{1}}(\b{r}_2) \; \d\b{r}_{1} \d\b{r}_{2} \d\b{k}_{1} \d\b{k}_{2} 
\nonumber\\
&=& -\frac{1}{2N(2{\pi})^{6}} \iint_{V_{k_{\text{F}}}} \iint_{V}  w_{\text{ee}}(r_{12}) ~ e^{-i\b{k}_{12}\cdot\b{r}_{12}}\; \frac{E_{k_{1}}E_{k_{2}}+(\b{k}_{1}\cdot\b{k}_{2})c^{2}+c^{4}}{E_{k_{1}}E_{k_{2}}} \;  \d\b{r}_{1} \d\b{r}_{2} \d\b{k}_{1} \d\b{k}_{2}
\nonumber\\
&=&-\frac{1}{2n(2{\pi})^{6}} \iint_{V_{k_{\text{F}}}} \tilde{w}_{\text{ee}}(k_{12}) \; \frac{E_{k_{1}}E_{k_{2}}+(\b{k}_{1}\cdot\b{k}_{2})c^{2}+c^{4}}{E_{k_{1}}E_{k_{2}}} \; \d\b{k}_{1} \d\b{k}_{2} 
\nonumber\\
	&=&-\frac{3~k_{\text{F}}}{4{\pi}}\Bigg(\frac{5}{6} + \frac{1}{3}{\ct}^{2} + \frac{2}{3} \sqrt{1+ {\ct}^{2}} ~ \text{arcsinh}\left(\frac{1}{\ct}\right) - \frac{1}{3}\Bigg(1+ {\ct}^{2}\Bigg)^{2}~\text{ln}\left(1+\frac{1}{{\ct}^{2}}\right) -\frac{1}{2}\left(\sqrt{1+ {\ct}^{2}} -{\ct}^{2} \text{arcsinh}\left(\frac{1}{\ct}\right)\right)^{2} ~\Bigg),
\label{epsxC}
\end{eqnarray}
where we have introduced $\b{k}_{12} = \b{k}_1 - \b{k}_2$, $k_{12}=|\b{k}_{12}|$, $\tilde{w}_{\text{ee}}(k_{12}) = 4{\pi}/k_{12}^{2}$ which is the Fourier transform of the Coulomb interaction potential, and $\ct = {c}/{k_{\text{F}}}$. The final expression in Eq.~(\ref{epsxC}) corresponds to the expression given in Refs.~\onlinecite{Jan-NC-62,MacVos-JPC-79}. The Breit exchange energy per particle of the RHEG can be calculated in a similar way
\begin{multline}
~~~~~~~~~~~~~~~~~~~ {\epsilon}_{\text{x}}^{\text{B}} = \frac{1}{4N} \frac{V^{2}}{(2{\pi})^{6}} \iint_{V_{k_{\text{F}}}} \iint_{V}  \sum_{{\sigma}_{1},{\sigma}_{2}={\downarrow},{\uparrow}} ~w_{\text{ee}}(r_{12}) \Bigg({\psi}_{\b{k}_{1},{\sigma}_{1}}^{\dagger}(\b{r}_1)\bm{\alpha}_{1}{\psi}_{\b{k}_{2},{\sigma}_{2}}(\b{r}_1)~\cdot~{\psi}_{\b{k}_{2},{\sigma}_{2}}^{\dagger}(\b{r}_2)\bm{\alpha}_{2}{\psi}_{\b{k}_{1},{\sigma}_{1}}(\b{r}_2)
\\
~~~~~~~~~~~~~~~~~ + \frac{{\psi}_{\b{k}_{1},{\sigma}_{1}}^{\dagger}(\b{r}_1)(\bm{\alpha}_{1}\cdot\b{r}_{12}){\psi}_{\b{k}_{2},{\sigma}_{2}}(\b{r}_1)~{\psi}_{\b{k}_{2},{\sigma}_{2}}^{\dagger}(\b{r}_2)(\bm{\alpha}_{2}\cdot\b{r}_{12}){\psi}_{\b{k}_{1},{\sigma}_{1}}(\b{r}_2)}{r_{12}^{2}}\Bigg) \d\b{r}_{1} \d\b{r}_{2}  \d\b{k}_{1} \d\b{k}_{2}
\\
= \frac{1}{2N(2{\pi})^{6}}\iint_{V_{k_{\text{F}}}} \iint_{V} w_{\text{ee}}(r_{12}) ~ e^{-i\b{k}_{12}\cdot\b{r}_{12}} ~\frac{c^{2}}{E_{k_{1}}E_{k_{2}}} \Bigg( \frac{E_{k_{2}}+c^{2}}{E_{k_{1}}+c^{2}} k_{1}^{2}  + \frac{E_{k_{1}}+c^{2}}{E_{k_{2}}+c^{2}} k_{2}^{2}  \Bigg)  \d\b{r}_{1} \d\b{r}_{2}  \d\b{k}_{1} \d\b{k}_{2} ~~~~~~~~~~~~
\\
= \frac{1}{2n(2{\pi})^{6}} \iint_{V_{k_{\text{F}}}} \tilde{w}_{\text{ee}}(k_{12}) \frac{c^{2}}{E_{k_{1}}E_{k_{2}}} \Bigg( \frac{E_{k_{2}}+c^{2}}{E_{k_{1}}+c^{2}}k_{1}^{2} + \frac{E_{k_{1}}+c^{2}}{E_{k_{2}}+c^{2}} k_{2}^{2} 
\Bigg)\d\b{k}_{1} \d\b{k}_{2} ~~~~~~~~~~~~~~~~~~~~~~~~~~~~~~~~~~~~~~~~~~~~~
\\
	= \frac{3~k_{\text{F}}}{4{\pi}}\Bigg(1 - 2\Big(1+{\ct}^{2}\Big)\Bigg(1 - {\ct}^{2} \Bigg(-2~\text{ln}\left({\ct}\right) + \text{ln} \left( 1 + {\ct}^{2} \right) \Bigg)\Bigg) + 2\left(\sqrt{1+ {\ct}^{2}} -{\ct}^{2} \text{arcsinh}\left(\frac{1}{\ct}\right) \right)^{2} \Bigg) ~,~~~~~~~~~~~~~~~~~~~~~~~~ 
\label{epsxB}
\end{multline}
which corresponds to the expression given in Ref.~\onlinecite{RamRaj-PRA-82}. The sums over spins in Eqs.~(\ref{epsxC}) and (\ref{epsxB}) are explicitly calculated in the Supplementary Information~\cite{PaqTou-JJJ-XX-note}.

\subsection{Effective Coulomb-Breit exchange hole}

The Coulomb exchange hole can be written as an integral over the wave vectors
\begin{eqnarray}
&& n_{\text{x}}^{\text{C}}(r_{12}) 
\nonumber\\
&=& - \frac{1}{(2{\pi})^{6}n} \iint_{V_{k_{\text{F}}}} e^{-i\b{k}_{12} \cdot\b{r}_{12}} \; \frac{E_{k_{1}}E_{k_{2}}+(\b{k}_{1}\cdot\b{k}_{2})c^{2}+c^{4}}{E_{k_{1}}E_{k_{2}}} \;  \d\b{k}_{1} \d\b{k}_{2} ~~~~~~~~~~~~~~~~~~~~~~~~~~~~~~~~~~~~~~~~~~~~~~~~~~~~~~~~~~~~~~~~~~~~~~~~~~~~~~~~~~~~~~~~~~~~~~ 
\nonumber\\
&=& - \frac{2}{16{\pi}^{4}n}\Bigg( ~\Bigg[\int_{0}^{k_{\text{F}}}k^{2}~j_{0}(kr_{12})\Big(1+\frac{c}{\sqrt{k^{2}+c^{2}}}\Big) \d k\Bigg]^{2} + \Bigg[\int_{0}^{k_{\text{F}}}k^{2}~j_{0}(k r_{12})\Big(1-\frac{c}{\sqrt{k^{2}+c^{2}}}\Big)\d k\Bigg]^{2} + 2 \Bigg[\int_{0}^{k_{\text{F}}} \frac{k^{3}}{\sqrt{k^{2}+c^{2}}}~j_{1}(k r_{12})\d k\Bigg]^{2} ~\Bigg), ~~~~~~~
\end{eqnarray}
where the last expression is obtained after integration over the angle coordinates, and $j_{\nu}$ are the spherical Bessel functions. After repeated integrations by parts, using for a general function $f$,
\begin{equation}
	\int_{0}^{k_{\text{F}}} k^{\nu+2} j_{\nu}(k r_{12}) f(k) dk 
	= \left[\frac{k^{\nu+2}}{r_{12}} j_{\nu+1}(k r_{12}) f(k)\right]_{0}^{k_{\text{F}}} - \int_{0}^{k_{\text{F}}}\frac{k^{\nu+2}}{r_{12}} j_{\nu+1}(k r_{12}) f'(k) dk,
\end{equation}
we obtain the expression of the Coulomb exchange hole as
\begin{equation}
	n_{\text{x}}^{\text{C}}(r_{12}) = - \frac{9}{4} n\frac{1}{(k_{\text{F}}r_{12})^{2}}\Bigg[j_{1}(k_{\text{F}} r_{12})^{2}+ (1-{\lambda})A_{\lambda}(k_{\text{F}} r_{12})^{2}
	+ {\lambda}B_{\lambda}(k_{\text{F}} r_{12})^{2}\Bigg],
\end{equation}
where $\lambda=1/(1+\ct^2)$ and
\begin{eqnarray}
	A_{\lambda}(k_{\text{F}} r_{12}) = \sum_{\nu=0}^{\infty} \frac{(2\nu+1)!!}{(2\nu+1)}~j_{\nu+1}(k_{\text{F}} r_{12})\left(\frac{\lambda}{k_{\text{F}} r_{12}} \right)^{\nu},
\nonumber\\
	B_{\lambda}(k_{\text{F}} r_{12}) = \sum_{\nu=0}^{\infty} \frac{(2\nu+1)!!}{(2\nu+1)}~j_{\nu+2}(k_{\text{F}} r_{12})\left(\frac{\lambda}{k_{\text{F}} r_{12}} \right)^{\nu}.
\end{eqnarray}
This is the expression given in Refs.~\onlinecite{Ell-JPB-77,MacVos-JPC-79}. Note that there are some typos in the expression given in Ref.~\onlinecite{Ell-JPB-77}. We now extend the previous derivation to the case of the Breit interaction. The associated exchange hole can be expressed as
\begin{eqnarray}
	n_{\text{x}}^{\text{B}}(r_{12}) &=& \frac{1}{(2{\pi})^{6}n}\iint_{V_{k_{\text{F}}}} ~ e^{-i\b{k}_{12}\cdot\b{r}_{12}} ~\frac{c^{2}}{E_{k_{1}}E_{k_{2}}} \Bigg(\frac{E_{k_{2}}+c^{2}}{E_{k_{1}}+c^{2}}k_{1}^{2} + \frac{E_{k_{1}}+c^{2}}{E_{k_{2}}+c^{2}} k_{2}^{2} \Bigg) \d\b{k}_{1} \d\b{k}_{2} 
\nonumber\\
&=& \frac{1}{2{\pi}^{4}n} \int_{0}^{k_{\text{F}}} k_{1}^{2} j_{0}(k_{1} r_{12})\Big( 1-\frac{c}{\sqrt{k_{1}^{2}+c^{2}}}\Big) \d k_{1} \times \int_{0}^{k_{\text{F}}} k_{2}^{2} j_{0}(k_{2} r_{12})\Big( 1+\frac{c}{\sqrt{k_{2}^{2}+c^{2}}}\Big) \d k_{2}, 
\end{eqnarray}
which, after using the same integration by parts as before, gives
\begin{equation}
	n_{\text{x}}^{\text{B}}(r_{12}) = -\frac{9}{2}n \frac{1}{(k_{\text{F}} r_{12})^{2}}\Bigg[-j_{1}(k_{\text{F}} r_{12})^{2} + (1-{\lambda}) A_{\lambda}(k_{\text{F}} r_{12})^{2}\Bigg].
\end{equation}
\end{widetext}

\section{Pad\'e approximants}
\label{app:pade}

We recall here how to calculate Pad\'e approximants and check their convergence~\cite{BenOrs-BOOK-99}. Consider a function $F$ which is asymptotic to the following (divergent) power series, as $z\to 0$,
\begin{eqnarray}
F(z) \sim \sum_{i=0}^{\infty} {\alpha}_{i}z^{i}.
\label{powerseries}
\end{eqnarray}
Its Pad\'e approximant of order $(N,M)$ is
\begin{eqnarray}
	P_{M}^{N}(z) &=& \frac{\sum_{i=0}^{N}A_{i} z^{i}}{\sum_{i=0}^{M} B_{i} z^{i}},
\label{PMM}
\end{eqnarray}
where $B_{0} = 1$ without loss of generality, and the other $N+M+1$ coefficients $A_{i}$ and $B_{i}$ are determined so that the power expansion of Eq.~(\ref{PMM}) matches the power series of Eq.~(\ref{powerseries}) up to order $N+M$. This gives the following matrix equation for the determining the coefficients $B_{1},B_2,...,B_{M}$
\begin{eqnarray} 
          \begin{pmatrix} {\alpha}_{N} & {\alpha}_{N-1} & ... & {\alpha}_{1} \\  {\alpha}_{N+1} & {\alpha}_{N} & ... & {\alpha}_{2} \\ \vdotswithin{M-3}  & \vdotswithin{M-3}  & \ddots & \vdotswithin{M-3} \\ {\alpha}_{N+M-1} & {\alpha}_{N+M-2} & ... & {\alpha}_{N} \end{pmatrix}
                  \begin{pmatrix} B_{1} \\ B_{2} \\ \vdotswithin{M-3} \\ B_{M} \end{pmatrix} = - \begin{pmatrix} {\alpha}_{N+1} \\ {\alpha}_{N+2} \\ \vdotswithin{M-3} \\ {\alpha}_{N+M} \end{pmatrix}, \;\;
\end{eqnarray}
and then the coefficients $A_0,A_{1},...,A_{N}$ are simply given by
\begin{eqnarray}
        A_{i} &=& \sum_{j=0}^{i} {\alpha}_{i-j}~B_{j}.
\end{eqnarray}

A lot about convergence of Pad\'e approximants is known for the special case where $F(z)$ is a so-called Stieltjes function. As prescribed in Ref.~\onlinecite{BenOrs-BOOK-99}, it can be checked whether $F(z)$ is a Stieltjes function by verifying the following four properties: (1) $F(z)$ is analytic in the cut complex plane $|\text{arg} \; z|<\pi$; (2) $\lim_{z\to\infty} F(z) =C$ where $C$ is a nonnegative real constant; (3) $F(z)$ has an asymptotic series representation of the sign-alternating form $\sum_{i=0}^{\infty} {a}_{i}(-z)^{i}$ where $a_i \geq0$; (4) $-F(z)$ is Herglotz, i.e. $\text{sgn}[\text{Im}(-F(z))] = \text{sgn}[\text{Im} \; z]$. If these four properties are satisfied, then it can shown that, for any real positive $z$,  the Pad\'e sequences $P_{M+1}^{M}(z)$ and $P_{M}^{M}(z)$ both converge as $M\to\infty$ and sandwich $F(z)$
\begin{eqnarray}
\lim_{M \to \infty} P_{M+1}^{M}(z) \leq F(z) \leq \lim_{M \to \infty} P_{M}^{M}(z). 
\end{eqnarray}
Obviously, if in addition the two limits are equal, then the two Pad\'e sequences converge to $F(z)$ as $M\to\infty$ for any real positive $z$. 

We have checked the convergence of the Pad\'e approximants in the case of the full-range Coulomb and Breit exchange energies per particle for which we have explicit exact expressions, Eqs.~(\ref{epsxCtext}) and~(\ref{epsxBtext}). For the Coulomb term, defining $F_{\text{x}}^{\text{C}}(z)=-{\epsilon}_{\text{x}}^{\text{C}}(\ct)$ with $z=1/\ct^2$, numerical investigations suggest that $F_{\text{x}}^{\text{C}}(z)$ satisfies properties (1) to (4) and is thus a Stieltjes function. Moreover, the Pad\'e approximants $P_{M+1}^{M}(z)$ and $P_{M}^{M}(z)$ appear numerically to converge to the unique limit $F_{\text{x}}^{\text{C}}(z)$ as $M\to\infty$ for real positive $z$. For the Breit term, defining $F_{\text{x}}^{\text{B}}(z)={\epsilon}_{\text{x}}^{\text{B}}(\ct)$ with $z=1/\ct^2$, numerical investigations suggest that $F_{\text{x}}^{\text{B}}(z)$ satisfies properties (1) to (3) but not (4), and thus is not a Stieltjes function. Nevertheless, the Pad\'e approximants $P_{M+1}^{M}(z)$ and $P_{M}^{M}(z)$ still appear numerically to converge to the unique limit $F_{\text{x}}^{\text{B}}(z)$ as $M\to\infty$ for real positive $z$. This does not come as a surprise since it is known that Pad\'e approximants often converge for functions which are not Stieltjes. We expect the same convergence properties for the Pad\'e approximants in the case of the short-range Coulomb and Breit interactions.


\end{document}